\documentclass[twocolumn,pre]{revtex4-1}

\usepackage{times}
\usepackage{amsmath}
\usepackage{amssymb}
\usepackage{graphicx}
\usepackage[unicode=true,bookmarks=false,breaklinks=false,pdfborder={0 0 1},backref=false,colorlinks=true]{hyperref}
\hypersetup{colorlinks,citecolor=blue,filecolor=black,linkcolor=red,urlcolor=blue}

\begin{document}

\title{Fractal scattering of Gaussian solitons in directional couplers with
logarithmic nonlinearities}

\author{Rafael M. P. Teixeira}
\address{Instituto de F\'{i}sica, Universidade Federal de Goi\'as, 74.690-900,
Goi\^ania, Goi\'as, Brazil}
\author{Wesley B. Cardoso}
\affiliation{Instituto de F\'{i}sica, Universidade Federal de Goi\'as, 74.690-900,
Goi\^ania, Goi\'as, Brazil}

\begin{abstract}
In this paper we study the interaction of Gaussian solitons in a dispersive
and nonlinear media with log-law nonlinearity. The model is described
by the coupled logarithmic nonlinear Schr\"odinger equations, which
is a nonintegrable system that allows the observation of a very rich
scenario in the collision patterns. By employing a variational approach
and direct numerical simulations, we observe a fractal-scattering
phenomenon from the exit velocities of each soliton as a function
of the input velocities. Furthermore, we introduce a linearization
model to identify the position of the reflection/transmission window
that emerges within the chaotic region. This enable us the possibility
of controlling the scattering of solitons as well as the lifetime
of bound states.
\end{abstract}

\maketitle

\section{Introduction \label{sec:Introduction}}

A soliton is a solitary wave that arises due to a perfect balance
between the dispersive and nonlinear effects present in the system,
it maintains its shape when moving at constant speed or even when
it emerges from the interaction with another soliton (except for a
phase shift) \cite{Zabusky_PRL65}. Solitonic solutions have been
observed in various contexts, such as, in Bose-Einstein condensates
(BECs) \cite{Khaykovich_SCI02,Strecker_NAT02,Cornish_PRL06,Marchant_NC13,Burger_PRL99},
water waves \cite{Craig_PF06}, proteins \cite{Davydov_85}, DNA \cite{Yakushevich_04},
nonlinear fiber optics \cite{Agrawal_01,Hasegawa_95} as temporal
solitons and as spatial optical solitons in a cell filled with sodium
vapor \cite{Bjorkholm_PRL74}, liquid carbon disulphide \cite{Barthelemy_OC85},
photorefractive crystals \cite{Segev_PRL92}, semiconductor waveguides
\cite{Aitchison_EL92}, nematic liquid-crystal planar cells \cite{Beeckman_OE04},
etc.

Some nonlinear systems have its dynamics dictated by the well-known
nonlinear Schr\"odinger (NLS) equation, e.g., BECs and nonlinear fiber
optics \cite{Kivshar_03}. In some particular cases the NLS equation
appears as an integrable equation, i.e., it can be integrated exactly
by the inverse scattering transform method \cite{Zakharov_JETP72}.
Thus, solitary wave solutions behave like solitons, with the characteristics
described above. In a more complex scenario, when the NLS equation
presents nonintegrability, collision of solitary waves can show a
complex structure since the collision outcome can depend on the initial
conditions, presenting a fractal pattern \cite{Yang_PRL00,Tan_PRE01,Dmitriev_CHAOS02,Zhu_PRE07,Zhu_PRL08,Zhu_PD08,Zhu_SAM09,Hause_PRA10}.
Fractal structures in solitons' collisions are also reported in systems
described by other equations, such as, in the $\varphi^{4}$ model
\cite{Goodman_CHAOS08,Goodman_CHAOS15}, the sine-Gordon model \cite{Fukushima_PLA95,Higuchi_CSF98,Dmitriev_PRE01,Dmitriev_PB02,Dmitriev_PRE08},
etc.

In case of systems governed by coupled NLS equations the conditions
of integrability can also be attained, as is the case of Manakov equations,
which describes the interaction of two light waves at different wavelengths
copropagating along one of the principal axes of a birefringent single-mode
fiber \cite{Agrawal_01} or a two-component BEC with two-body interaction
and in absence of external potentials \cite{Myatt_PRL97,Stamper-Kurn_PRL98}.
Once again, a rich scenario arises when the system becomes nonintegrable
due to simple changes in the values of the couplings, the inclusion
of inhomogeneous terms or high-order terms in the coupled NLS equation.
In this case we have the possibility to get modulated localized solutions
\cite{Cardoso_PRE12,Cardoso_PLA10-2} or observe its fractal scattering
\cite{Yang_PRL00}. Another possibility to obtain this type of nonintegrable
systems is the use of directional couplers \cite{Kogelnik_IEEE76,Bergh_EL80,Streltsov_OL01,Alves_arXiv15},
which is composed by fibers that are generally twisted together and
then spot fused under tension such that the fused section is elongated
to form a biconical tapered structure, being based on the transfer
of energy by surface interaction between the fibers. The amount of
power taken from the main fiber or given to the it depends on the
length of the fused section of the fiber and the distance between
the cores of the fused fibers \cite{Biswas_06}. 

In several distinct scenarios in physics and in other areas of nonlinear
science the systems under consideration are well described by NLS
equations with logarithm nonlinearity, as for example, in dissipative
systems \cite{Hernandez_PA81}, in nuclear physics \cite{Hefter_PRA85},
in optics \cite{Krolikowski_PRE00,Buljan_PRE03}, capillary fluids
\cite{DeMartino_WASCOM04}, and even in magma transport \cite{Martino_EPL03}.
Also, the study of localized Gaussian-shaped solutions (Gaussons)
was central point in Ref. \cite{Biswas_CNSNS10}. In Refs. \cite{Biswas_OLT12,Zhou_Optik13}
were provided a set of exact optical soliton solutions of the NLS
equation with Kerr and non-Kerr (including logarithm) nonlinearities
and in presence of different perturbations. A wavelet formulation
was proposed in Ref. \cite{Hilal_Optik14}, which is suitable for
analyzing the optical soliton signals by introducing a nonlinear wavelet-like
basis of scaling functions made by localized analytical nonlinear
solutions. The modulation of localized solutions in inhomogeneous
NLS equations with logarithm nonlinearity was studied in Ref. \cite{Calaca_CNSNS14}
and in a system with time-dependent dispersion and nonlinearity in
\cite{Biswas_AMC10}. Quasi-stationary optical solitons in non-Kerr
media in presence of high-order terms was investigated in Ref. \cite{BISWAS_JNOPM11}.

In the present paper we study the fractal scattering of solitons'
collisions in a directional coupler in presence of logarithm nonlinearities.
To this end, we apply the variational approach by assuming Gaussian
symmetric solutions for each branch of the directional coupler to
construct a reduced ordinary differential equations (ODE) model \cite{Yang_10}.
This model allows us to analytically investigate the formation of
fractal patterns and the properties of scattered solitons. Within
the reflection windows, we can distinguish the conditions for the
shape-oscillations obtained by the solitons during its interaction,
which is due to the exchange of energy between the oscillation and
propagation modes of the solutions. We also employ direct numerical
simulations to confirm the applicability of the ODE model.

This paper is organized as follows. In the next section we present
the theoretical model and the analytical result obtained by the variational
approach. Numerical results for the reduced ODE model and direct numerical
simulations are presented in Sec. \ref{sec:Numerical-Results}. Particularly,
we present in Subsec. \ref{sub:Analysis-of-the} the analysis of the
dynamics of solitons' collisions from both approaches. Concluding
remarks are given in Sec. \ref{sec:Conclusion}.

\section{Theoretical Model \label{sec:Theoretical-Model}}

For twin-core couplers in a system with log-law nonlinearity, the
wave propagation at relatively high field intensities is described
by coupled nonlinear equations. In the dimensionless form, they are
given by \cite{Biswas_06}\begin{subequations}
\begin{eqnarray}
i\phi_{z} & = & -\dfrac{1}{2}\phi_{TT}+g\ln(|\phi|^{2})\phi+\Gamma\psi,\label{cnse1}\\
i\psi_{z} & = & -\dfrac{1}{2}\psi_{TT}+g\ln(|\psi|^{2})\psi+\Gamma\phi,\label{cnse2}
\end{eqnarray}
\end{subequations}where $z$ is the longitudinal coordinate and $T=t-\beta z$,
with time $t$, is the retarded time moving with the group velocity
of the fundamental mode in one of the waveguide with propagation constant
$\beta$, the other mode is assumed to be the same value of the propagation
constant. $\psi=\psi(z,T)$ and $\phi=\phi(z,T)$ are complex amplitudes
of the wave envelopes in the respective cores of the optical fibers,
$g$ is a negative coefficient providing a self-focusing nonlinearity,
and $\Gamma$ is the coefficient that binds the two pulses propagating
through these cores. When there is no coupling, i.e., $\Gamma=0$,
one can easily found a localized solution for both decoupled fields
by assuming (and similarly for $\psi(z,T)$) 
\begin{equation}
\phi(z,T)=U(T)e^{i\mu z},
\end{equation}
which transforms the Eq. \eqref{cnse1} in an ODE that yields the
solution 
\begin{equation}
\phi(z,T)=\exp\left[g(T-T_{0})^{2}+\frac{1}{2}\left(1-\frac{\mu}{g}\right)+i\mu z\right],\label{Gausson}
\end{equation}

\noindent where $\mu$ is an arbitrary real constant and $T_{0}$
is the Gaussian peak position at $z=0$. Note that the solution (\ref{Gausson}),
since $g<0$, have a Gaussian-shaped profile. Thus, in the literature
these solutions are also known as \emph{Gaussons}.

\subsection{Variational approach \label{sub:Variational-approach}}

The equations of motion \eqref{cnse1} and \eqref{cnse2} comes from
the following Lagrangian density 
\begin{eqnarray}
\mathcal{L} & = & \dfrac{i}{2}(\psi\psi_{z}^{*}-\psi^{*}\psi_{z})+\dfrac{i}{2}(\phi\phi_{z}^{*}-\phi^{*}\phi_{z})\nonumber \\
 & - & \dfrac{1}{2}|\psi_{T}|^{2}-\dfrac{1}{2}|\phi_{T}|^{2}-\Gamma(\psi\phi^{*}+\psi^{*}\phi)\nonumber \\
 & - & g|\psi|^{2}[\ln(|\psi|^{2})-1]-g|\phi|^{2}[\ln(|\phi|^{2})-1],\label{dens}
\end{eqnarray}
where the subscript ``$*$'' stands for complex conjugation.

Analytical results can be obtained through a variational approach
that uses a functional form (ansatz) for fields $\psi$ and $\phi$.
To this end, we use an ansatz that reproduces very well most of the
features involving interactions of solitons and provide an exact solution
when $\Gamma\rightarrow0$. Assuming the fields to be symmetric with
respect to $T=0$, we consider the ansatz given by\begin{subequations}
\begin{align}
\psi(z,T) & =\eta e^{\left\{ g\left(\frac{T-\xi}{w}\right)^{2}+i\left[\frac{v}{4}(T-\xi)+\frac{b}{2w}(T-\xi)^{2}+\sigma\right]\right\} },\label{sol1}\\
\phi(z,T) & =\eta e^{\left\{ g\left(\frac{T+\xi}{w}\right)^{2}+i\left[\frac{-v}{4}(T+\xi)+\frac{b}{2w}(T+\xi)^{2}+\sigma\right]\right\} },\label{sol2}
\end{align}
\end{subequations}where the dimensionless variational parameters
$\eta$, $w$, $v$, $b$, $\xi$ and $\sigma$ are $z$-dependent
functions and represents the amplitude, width, velocity, chirp, position,
and phase, respectively, of both solitary waves. The chirp parameter
is responsible to induce shape-oscillations, which are seen as oscillations
of the amplitude and the width of the Gaussian profile of both solitary
waves. So, substituting Eqs. \eqref{sol1}-\eqref{sol2} into the
Lagrangian density \eqref{dens}, one can calculate the associated
Lagrangian by $L=\int_{-\infty}^{\infty}{\mathcal{L}}\,dT,$ which
gives, after performing the necessary integrations and some algebra,
the following result: 
\begin{eqnarray}
L & = & \eta^{2}w\sqrt{\dfrac{\pi}{2}}\left\{ \dfrac{\xi_{t}v}{2}-\dfrac{w^{2}}{2}\left(\dfrac{b_{t}}{2w}-\dfrac{bw_{t}}{2w^{2}}\right)\right.\nonumber \\
 & - & \dfrac{v^{2}}{16}-\dfrac{b^{2}}{4}-\dfrac{1}{w^{2}}+g\left(3-4\ln\eta\right)-\nonumber \\
 & - & \left.2\Gamma\exp\left[-\dfrac{\left(\frac{v}{2}-\frac{2b\xi}{w}\right)^{2}w^{4}+16\xi^{2}}{8w^{2}}\right]\right\} \,.\label{lagragian}
\end{eqnarray}

The equations of motion are the Euler-Lagrange equations derived from
\eqref{lagragian}. Since $\sigma$ is assumed to be a constant phase
factor, there are five equations, given by\begin{subequations}
\begin{eqnarray}
\eta^{2}w & = & K=\text{constant}\,,\label{euler-lag0}\\
\frac{d\xi}{dz} & = & \frac{v}{4}+4\Gamma\frac{\partial G}{\partial v},\label{euler-lag1}\\
\frac{dv}{dz} & = & -4\Gamma\frac{\partial G}{\partial\xi},\\
\frac{dw}{dz} & = & b+4\Gamma\frac{\partial G}{\partial b},\\
\frac{db}{dz} & = & \dfrac{4}{w^{3}}+\frac{4g}{w}-4\Gamma\frac{\partial G}{\partial w},\label{euler-lag}
\end{eqnarray}
\end{subequations}

\noindent where the function of variational parameters $G=G(\xi,w,v,b)$
is written as 
\begin{equation}
G=\exp\left[-\dfrac{\left(\frac{v}{2}-\frac{2b\xi}{w}\right)^{2}w^{4}+16\xi^{2}}{8w^{2}}\right]\,.
\end{equation}

In the case of decoupled fields ($\Gamma=0$ in Eqs. \eqref{cnse1}
and \eqref{cnse2}) we expect the individual norm conservation, i.e.,
$\mathcal{N}_{\psi}=\int_{-\infty}^{\infty}|\psi(z,T)|^{2}dT$ and
$\mathcal{N}_{\phi}=\int_{-\infty}^{\infty}|\phi(z,T)|^{2}dT$. In
the general case we attempt to the total norm, given by $\mathcal{N}=\mathcal{N}_{\phi}+\mathcal{N}_{\psi}$.
Also, the ansatz employed does not contain degrees of freedom that
would be responsible for radiation loss hence the total norm of both
solitary waves must be a conserved quantity of the system. In fact,
since the interaction is assumed to be symmetric, the norm is conserved
for each soliton individually. The calculation of individual norm
for both fields in equations in \eqref{sol1}-\eqref{sol2} results
in

\begin{equation}
\mathcal{N}_{\phi,\psi}=\eta^{2}w\sqrt{\dfrac{\pi}{2|g|}},
\end{equation}

\noindent providing that $\mathcal{N}\propto K$ and the Eq. \eqref{euler-lag0}
is attained from the norm conservation.

In order to investigate symmetric solitons' collisions, the system
of coupled ODE \eqref{euler-lag1}-\eqref{euler-lag} is solved numerically
subjected to the condition $\eta^{2}w=K$. For this we use $w_{0}=1$,
$b_{0}=0$, and $K=1$, for simplicity. Since the width oscillations
are induced by the chirp parameter, null initial values for it (i.e.,
$b=0$) guarantee that no shape-oscillations exist initially. In the
next section we show the results of our numerical simulations.

\section{Numerical Results \label{sec:Numerical-Results}}

\subsection{Reduced ODE model \label{sub:Reduced-ODE-model}}

\begin{figure}[b]
\centering \includegraphics[width=1\columnwidth]{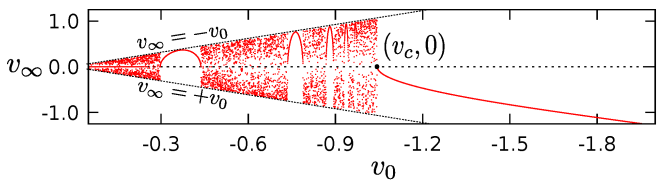}
\caption{(Color online) Exit velocity $v_{\infty}$ versus the input velocity
$v_{0}$ in the soliton interaction. The critical point $(v_{c},0)$
separates the region where the collision dynamics is irregular from
the region where the solitons collide elastically $(|v_{0}|>|v_{c}|)$.
Two straight lines given by $v_{\infty}=\pm v_{0}$ define a cone
in which all points are contained.}
\label{F1} 
\end{figure}

The numerical simulations of the coupled equations from the reduced
ODE model were performed using the $4^{\text{th}}$-order Runge-Kutta
method. We chose as initial conditions $\xi_{0}=10$, $w_{0}=1$,
$b_{0}=0,$ $\sigma=0$, and $K=1$. The initial separation between
the soltions is then $20$ units wide, which is found to guarantee
a negligible overlap of the wave packets at $z=0$. Here, we assume
the nonlinear and coupling coefficients given by $g=-1$ and $\Gamma=-0.2$,
respectively. Then, we study the scattering in soliton symmetric collisions
by using the initial velocity $v_{0}$ as the control parameter. In
the simulations we set the $z$-step value as $0.001$ and the $T$-range
in a symmetric interval with $100$ units wide. The program was developed
using the Fortran 95 language, in which we used double precision for
both real and complex numbers.

The solitons scattering simulations were executed over various intervals
of $v_{0}$, each with at least $5000$ points that correspond to
the same number of individual simulations with a fixed $v_{0}$, in
which we get the exit velocity ($v_{\infty}$) of both solitons at
the end of the interaction. We chose the soliton in the right position
as reference when recording the scattering data; its initial velocity
must be always negative in order to obtain a collisional scenario,
while for the soliton in the left position stands the opposite. The
exit velocity will assume negative (positive) values when the right
soliton exits toward the $T<0$ ($T>0$) region. In this sense, the
exit velocity is attained when the solitons are far apart from each
other, enough to ensure that no more interaction will occur in the
unbounded medium (\textcolor{black}{in our case we verified that it
should be about $20$ units wide in $T$}). The solitons can interact
for a very long time, forming a bound-state. So, we set a stopping
trigger to not let the program exceed a certain interaction time,
in our case it was achieved by setting $z_{max}=400$. When this maximum
is reached we assign $v_{\infty}$ as zero, i.e., we consider it as
a bound-state. 

\begin{figure}[tb]
\begin{centering}
\centering \includegraphics[width=1\columnwidth]{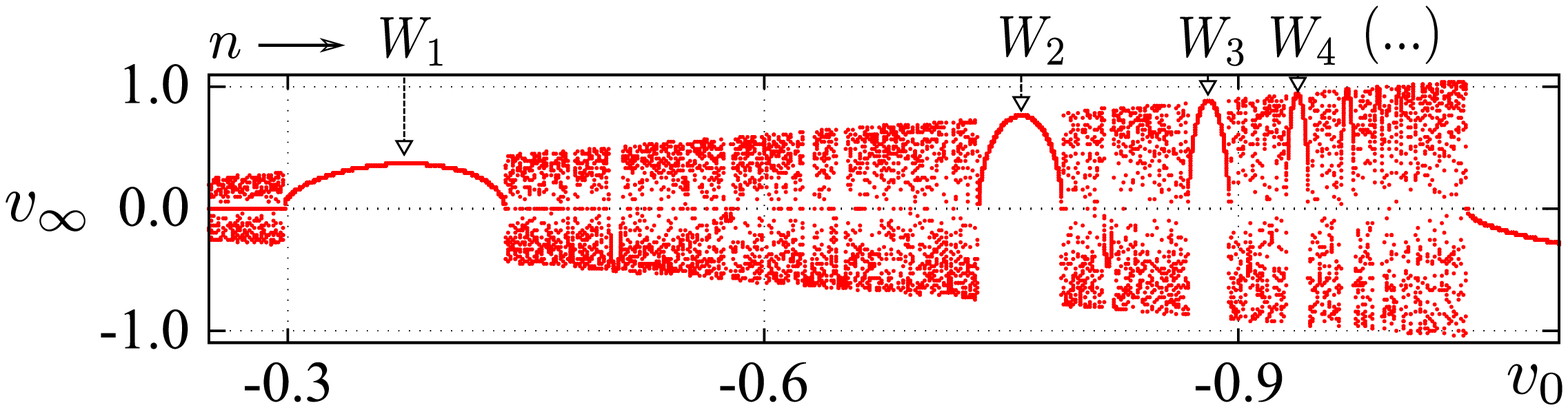} 
\includegraphics[width=0.8\columnwidth]{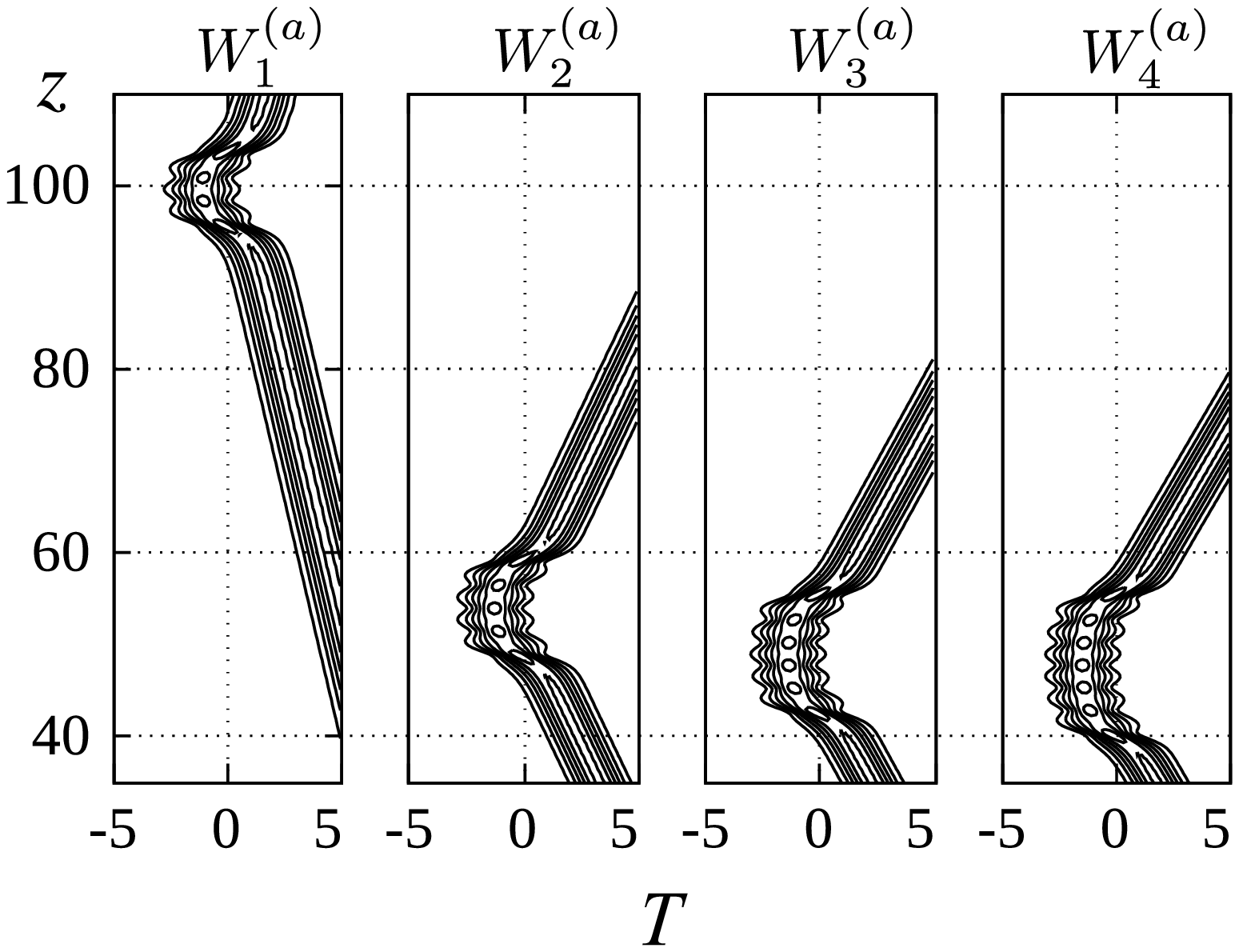} 
\par\end{centering}

\caption{(Color online) (Top) The exit versus input velocities in the soliton
scattering. The reflection windows $W_{n}$ are highlighted to evidence
the adopted sequence for $n$ (there is no windows before $W_{1}$).
(Bottom) Soliton profile ($|\psi(z,T)|^{2}$) during the interaction
for values of $v_{0}$ in the crest of the reflection windows with
$n=1,2,3$ and $4$. The crests are very close to the cone of maximum
exit velocity. The profile for the $|\phi(z,T)|^{2}$ is a mirrored
image of $|\psi(z,T)|^{2}$. }
\label{F2} 
\end{figure}

In the first simulation we use a $v_{0}$-interval starting at a relatively
low velocity ($>-0.05$) and ending at very large velocity ($<-5.0$).
We verified that for large values of $|v_{0}|$ the collision is elastic
and consists of only one interaction with the solitons passing through
each other. At lower velocities there is a critical point $(v_{c},0)$,
such that as $v_{0}$ approximates the critical velocity $v_{c}\approx-1.04$
the collision becomes each time more inelastic. If $v_{0}>v_{c}$
(i.e, $|v_{0}|<|v_{c}|$) the dynamics of the interaction changes
completely and the solitons collide in an unpredictable fashion. This
is shown in Fig. \ref{F1} for the exit velocity limited to the range
$[-0.05,-2.00]$. The irregular scattering of the solitons is characterized
by the high sensitivity of the initial condition, i.e., the collisional
velocity $v_{0}$. Note that all points in the plot are contained
in a cone of maximum exit velocity, which is given by the dotted-lines
($v_{\infty}=\pm v_{0}$), which means that in the current configuration
of our system the inequality $|v_{\infty}|\leq|v_{0}|$ holds, and
the solitons can not gain momentum through symmetric collisions.

The region $v_{0}>v_{c}$ is shown in more details in Fig. \ref{F2}
(top), where we detect the existence of several windows of lump-like
shape that repeats indefinitely as $v_{0}$ tends to $v_{c}$. In
these windows the soliton scattering is not sensitive to $v_{0}$,
in fact, we verified that inside each window the collision resembles
a reflection process, with the solitons forming a bound-state with
fixed lifetime ($z$-interval of existence of the bound-state) and
escaping toward the same region ($T>0$ or $T<0$), i.e., its initial
position. This type of window is called reflection window. An interesting
feature of the reflection window is that for any value of the initial
velocity within it, the soliton profile oscillates the same number
of times, in which one shape-oscillation was taken to be one period
of oscillation of the width of the solitons. This last statement leads
to the property of a fixed lifetime of the bound-state of the solitons,
as mentioned before. Also, the differences between the lifetimes of
bound-states of any two successive windows are always one shape-oscillation
period. We designate these windows by $W_{n}$, as seen in Fig. \ref{F2}
(top), with $n$ being the window index. The numbering starts with
the largest window ($n=1$). Also, in Fig. \ref{F2} (bottom) the
details of the collisions are shown for values of $v_{0}$ within
four successive reflection windows, the number of shape-oscillations
($N_{SO}^{n}$) was found to be related to $n$ by $N_{SO}^{n}=n+3$.
The value of $N_{SO}^{n}$ is obtained by counting the number of amplitude
peaks in the soliton, which is easily seen in the contour plots. We
noticed that these windows form a structure, which consist of a repetition
of windows separated by intervals that become smaller as $v_{0}$
approaches $v_{c}$. The window shape is basically a lump with the
crest almost tangent to the cone of maximum $v_{\infty}$, it becomes
narrower as closer it is from the critical point.

\begin{figure}[tb]
\centering \includegraphics[width=1\columnwidth]{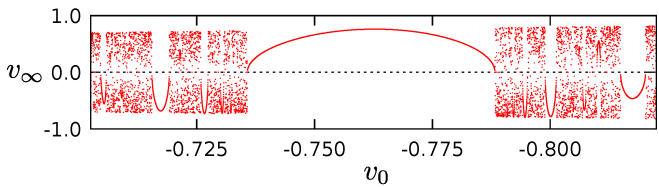} 
\caption{(Color online) Result for the simulation in an interval centered at
$W_{2}$ (Fig. \ref{F2} (top)). The size of the interval was chosen
to totally encompass the window structures that appear in the edges
of the reflection window $W_{2}$.}
\label{F3} 
\end{figure}

In Fig. \ref{F2} we noticed the existence of some smaller window
structures in the edges of each reflection window. Then, to provide
a better visualization of it we simulated in an interval centered
at $W_{2}$, where the resultant exit velocity graph is shown in Fig.
\ref{F3}. In this figure two structures appear clearly, one resembling
a mirrored image of the other. Besides, both are very similar to the
one in Fig. \ref{F2}, but with the windows being like valleys instead
of lumps. These are called transmission windows because the collisions
associated to it resemble a transmission process. As seen in the first
exit velocity graph, the windows become narrower and closer spaced
near the critical points. For the structures in Fig. \ref{F3} these
points are located in the edges of the reflection window $W_{2}$.

\begin{figure}[tb]
\centering \includegraphics[width=1\columnwidth]{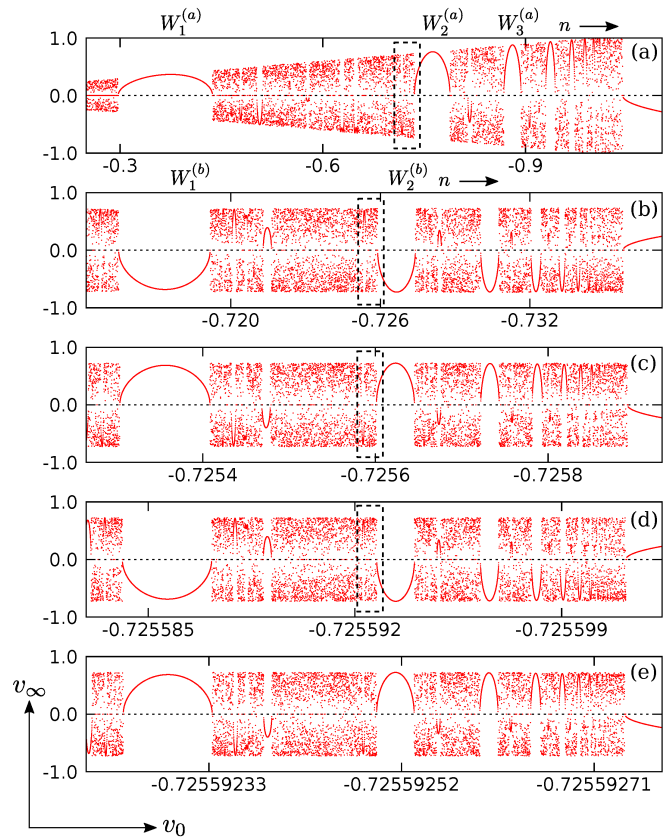}
\caption{(Color online) Results of successive amplifications of slim $v_{0}$-intervals
between certain reflection and transmissional windows as marked in
figures: (a) top part of Fig. \ref{F2} that is used to emphasize
the pattern repetition; (b)-(e) are successive amplifications of the
intervals marked in the exit velocity graphs, where we verify similar
window structures to those ones appearing in (a).}
\label{F4} 
\end{figure}

To explore the smaller structures embedded in the region $|v_{0}|<|v_{c}|$
we performed simulations amplifying the structures located in narrower
$v_{0}$-intervals near the edges of certain windows. Firstly, we
chose the left edge of the window $W_{2}$, which yielded a structure
of windows similar to the ones seen until now. Hence, we expected
that smaller structures with same pattern could be found by applying
the same procedure. So, we have adopted a protocol that consists of
choosing intervals for amplifications always in the left edge of every
second window of any structure that may appear in the successive amplifications.
Following this protocol we obtained the results shown in Fig. \ref{F4},
where the first plot is the same of Fig. \ref{F2} (top). The highlighted
intervals in the plots indicate the region of amplification, which
corresponds to the plot immediately bellow. Here we modified our notation
for labeling reflectional/transmissional windows, such that the superscript
$(s)$ were added to denote the structure in which the window is located,
i.e., $W_{n}^{(s)}$ is the label of the $n$-th window in the structure
in Fig. \ref{F4}(s), with $s=a,b,...,e$. We extend this notation
to the critical velocity associated with each structure, here denoted
by $v_{c}^{(s)}$.

We observed that for all scattering data obtained, zero exit velocities
have appeared only for collisions associated with initial velocities
between zero and the window $W_{1}^{(a)}$. In order to verify the
true outcome of these collisions, very long simulations were performed.
We found that the bound-state formed during the collision eventually
ends for a certain $z$ much larger than $z_{max}$. Thus there are
only two different scenarios concerning the solitons' collisions accordingly
to the reduced ODE model, namely, transmissional collision ($v_{\infty}<0$)
and reflectional collision ($v_{\infty}>0$). Our simulations show
that transmission and reflection windows are distributed along very
slim intervals within the chaotic region ($|v_{0}|<|v_{c}|$), where
the collision dynamic is indeed very sensitive to the choice of $v_{0}$.
The most interesting feature found for the exit velocities are the
structures that repeat embedded in itself, forming in the edges of
every window, as shown before in Fig. \ref{F3}. Besides, they have
a very similar pattern regarding the size and distribution of the
windows. It is clearly seem that Fig. \ref{F4}(a) resembles Figs.
\ref{F4}(c) and \ref{F4}(e), also the Fig. \ref{F4}(b) resembles
Fig. \ref{F4}(d). This suggests that the structures always appear
alternating the window type. Also, excepting in Fig. \ref{F4}(a),
they seem to be related by just one symmetry operation, i.e., a reflection
about the $v_{0}$-axis. 

By analyzing the window patterns for the structures in Fig. \ref{F4},
we found that the velocity $v_{n}^{(s)}$ in the crest (trough) of
a reflection (transmission) window $W_{n}^{(s)}$ and the critical
velocity $v_{c}^{(s)}$ of the structure given in \ref{F4}$(s)$
are very well related by \cite{Yang_10,Tan_PRE01}

\begin{equation}
\dfrac{1}{\sqrt{|(v_{c}^{(s)})^{2}-(v_{n}^{(s)})^{2}|}}=p^{(s)}n+q^{(s)}\,,\label{eq:linear1}
\end{equation}

\noindent where the coefficients $p^{(s)}$ and $q^{(s)}$ vary for
the different values of $s$. We illustrate this fact for the windows
until index $n=6$, the linear relation in \eqref{eq:linear1} is
satisfied with great precision for the Fig. \ref{F4}(a), while for
the other structures in Figs. \ref{F4}(b)-(e) we found relative small
standard deviations. These results allowed us to infer that the window
patterns in the intervals of our simulations satisfy a common relation
given by \eqref{eq:linear1}, however, it is not enough to conclude
that the pattern is closely the same, since the coefficients can not
be compared because the structures have different sizes. We solved
this problem by rescaling and displacing the structures, in a manner
that the left edge of $W_{1}^{(s)}$ is at the origin and the critical
point of every structure at $(-1,0)$. This transformation changes
the values of the initial velocities in the crest (trough), which
are denoted by $\widetilde{v}_{n}^{\,(s)}$. In this way we can say
that $\widetilde{v}_{c}^{\,(s)}\equiv-1$ for all $(s)$, with $\widetilde{v}_{n}^{\,(s)}<\widetilde{v}_{c}^{\,(s)}$.
So, the transformed structures have the same size. The coefficients
found for the first structure are $\widetilde{p}^{\,(a)}=0.325\pm0.006\,(1.85\%)$
and $\widetilde{q}^{\,(a)}=0.65\pm0.02\,(3.08\%)$, while for the
remaining structures we calculated the average of the coefficients
together with the standard deviations $\langle\widetilde{p}\rangle=0.280\pm0.003\,(1.24\%)$
and $\langle\widetilde{q}\rangle=0.662\pm0.004\,(0.68\%)$. Note that
these averages values have very small standard deviations, but there
is a relevant difference when comparing $\langle\widetilde{p}\rangle$
and $\widetilde{p}^{\,(a)}$. These results show that the window pattern
is almost the same for all structures after the first amplification.
We stress that the first exit velocity plot presents a subtle difference
among the others, which resides in the region where zero exit velocities
appeared (which is purely chaotic), differently from the other structures.
Hence a self-similarity argument can be applied only to Figs. \ref{F4}(b)-(e).

\begin{figure}[tb]
\begin{centering}
\centering \includegraphics[width=0.9\columnwidth]{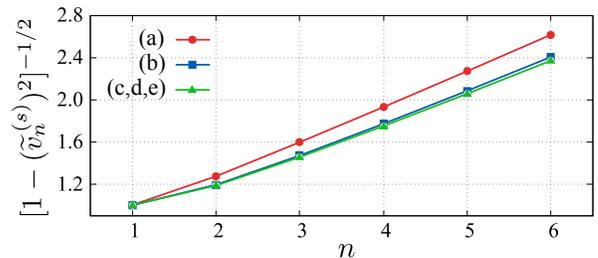} 
\par\end{centering}

\raggedright{}\caption{(Color online) Plot of $[1-(\widetilde{v}_{n}^{\,(s)})^{2}]^{-1/2}$
as a function of the the window index $n$. In curves (a)-(e) we used
data obtained by the structures shown in Fig. \ref{F4}(a)-(e), respectively.}
\label{FL1} 
\end{figure}

In Fig. \ref{FL1} we show the quantity $[1-(\widetilde{v}_{n}^{\,(s)})^{2}]^{-1/2}$
as a function of the window index $n$. Note that the curves related
to the structure presented in Figs. \ref{F4}(c)-(e) are too close
to be distinguished and the curve related to Figs. \ref{F4}(b) is
also very close from those. Also, our observations indicate that this
pattern prevails for any window structure embedded into the Fig. \ref{F4}(a),
which is a characteristic feature of a fractal pattern and demonstrate
the chaotic behavior of the scattering. Furthermore, we verify that
the size ($L^{(s)}$) of the window structure in Fig. \ref{F4}(s),
which is the length between the left edge of the first window and
the critical velocity $v_{c}^{(s)}$, nicely satisfies the formula

\begin{equation}
\log_{10}L^{(s)}=Rj+r\,,\label{eq:linear2}
\end{equation}

\noindent where $j=1$ for Fig. \ref{F4}(a), $j=2$ for Fig. \ref{F4}(b),
and so on. Moreover, the integer number $j-1$ is the number of amplifications.
In Eq. \eqref{eq:linear2} the linear fit yielded $R=-1.639\pm0.008\,(0.47\%)$
and $r=1.56\pm0.02\,(1.63\%)$. After some algebra, the length $L^{(s)}$
can be expressed in terms of the $j$ and $L^{(a)}$ as

\begin{equation}
L^{(s)}=10^{R(j-1)}L^{(a)}\,.\label{eq:length}
\end{equation}

\noindent Since $R$ is a negative constant, the factor that multiplies
$L^{(a)}$ in \eqref{eq:length} is less than unity and we call it
as reduction factor. Note that this results implies that length of
the window structures decreases exponentially with the number of amplifications.
Therefore the zoom ratio is closely the same at each amplification,
which is given by $10^{-R}\approx43.6$ (the reduction ratio is the
inverse). As an example, to emphasize how is this reduction, for structure
in Fig. \ref{F4}(e) ($j=5$) one obtains the reduction factor $\sim10^{-7}$,
while for structure in Fig. \ref{F4}(d) ($j=4$) it is $\sim10^{-2}$.

\begin{figure*}[tb]
\centering\includegraphics[clip,width=0.4\linewidth]{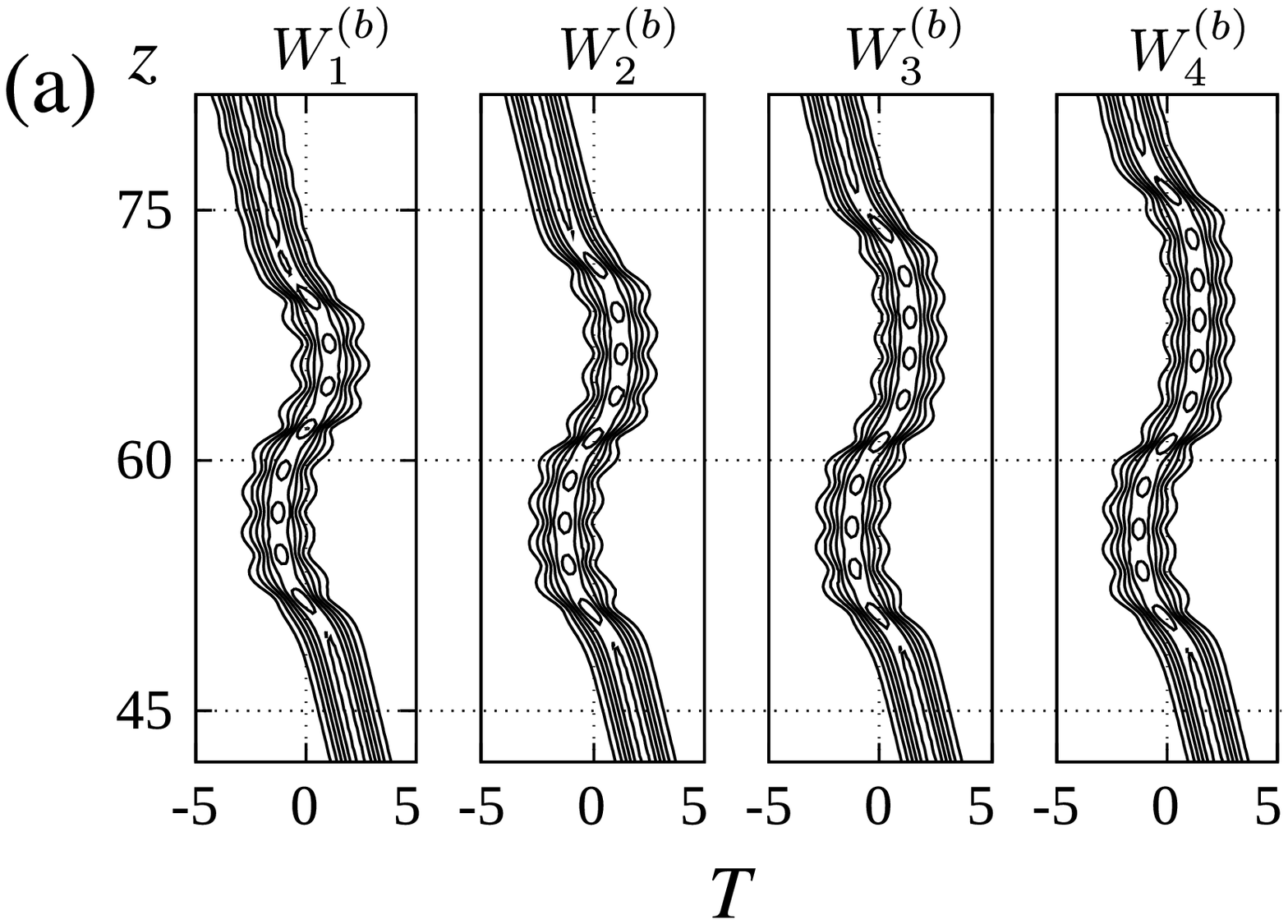}\hfil
\includegraphics[clip,width=0.4\linewidth]{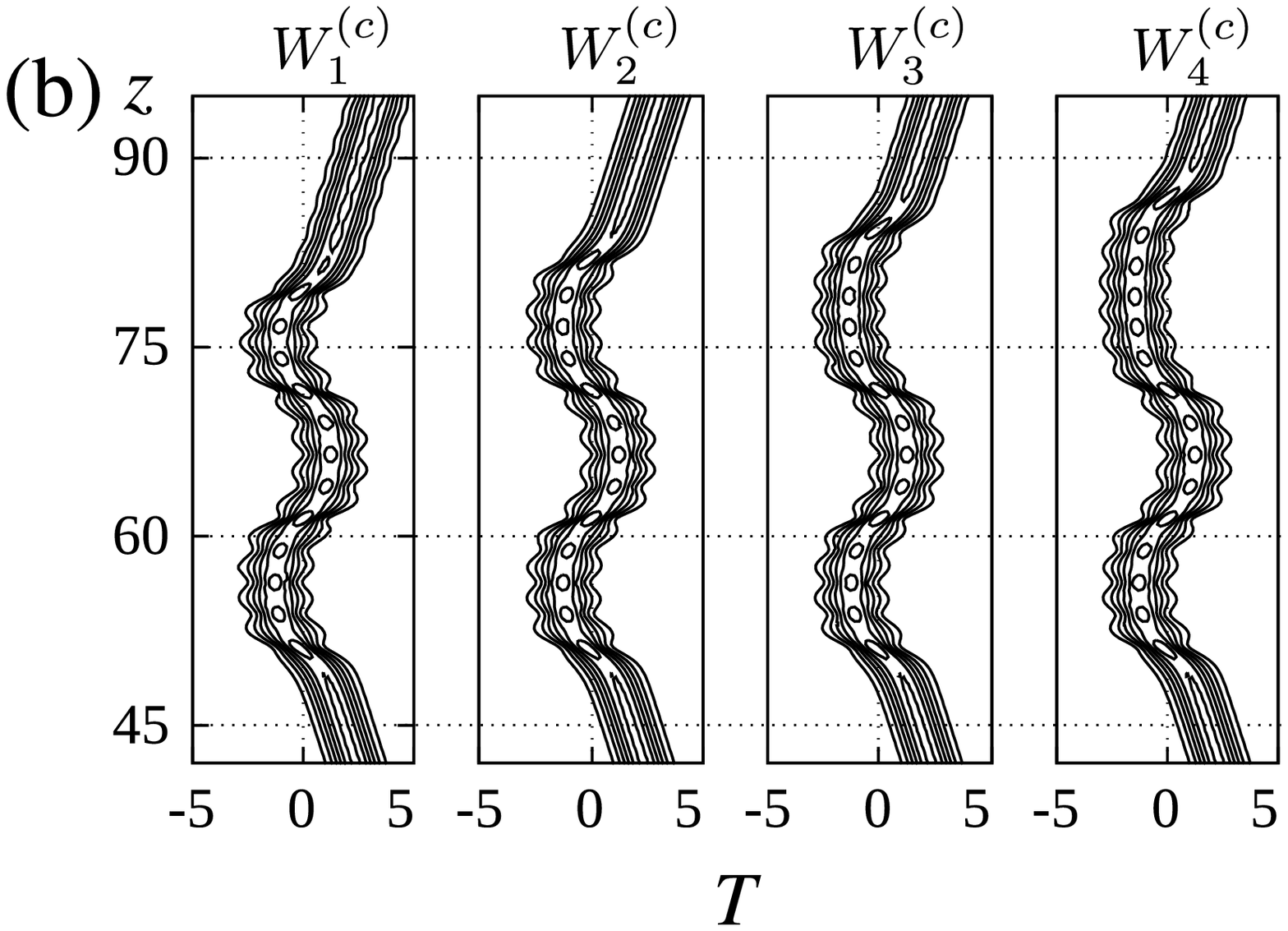} \includegraphics[clip,width=0.4\linewidth]{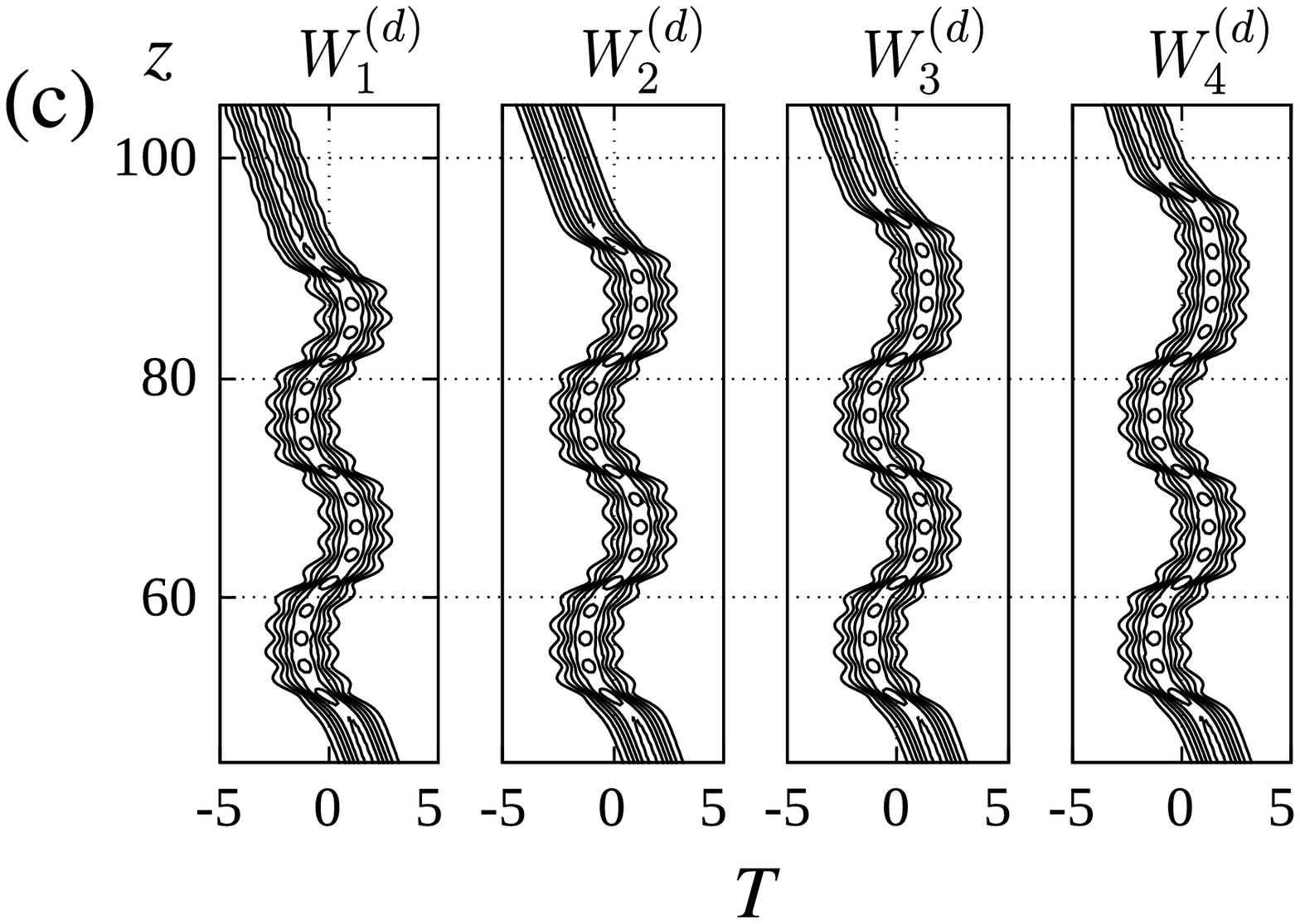}\hfil
\includegraphics[clip,width=0.4\linewidth]{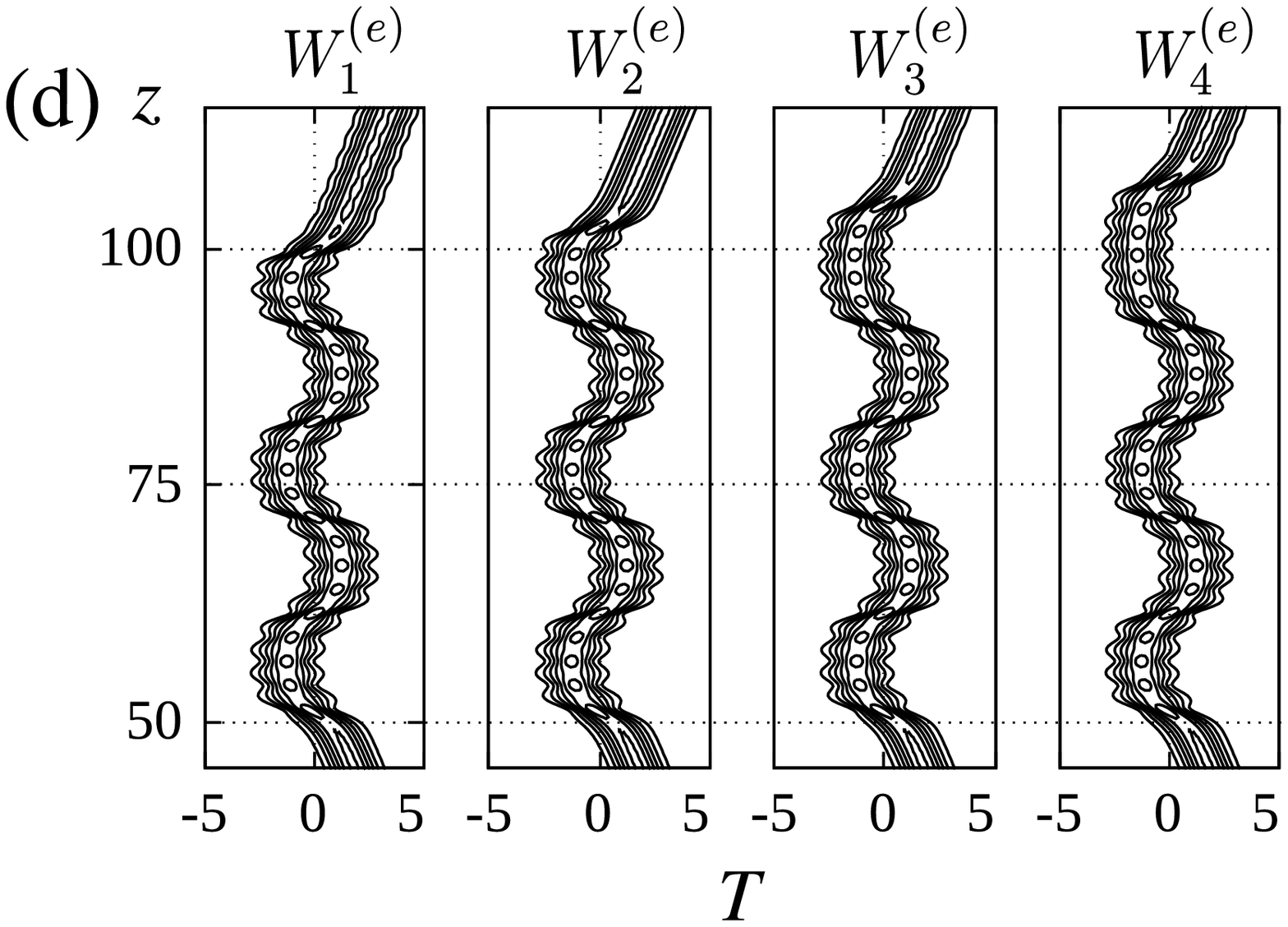}\caption{Profile of the localized solution $|\psi(z,T)|^{2}$ during the interaction
with the field $|\phi(z,T)|^{2}$ obtained using the reduced ODE model
of Eqs. \eqref{cnse1} and \eqref{cnse2}. The initial velocity values
are taken within four successive windows in the structure shown in
Fig. \ref{F4}(s), with $s=\{b,c,d,e\}$. The number of shape-oscillations
in the bound-state of the solitons yields $N_{SO}^{n}=(n-1)+m^{(s)}\times4+4$
for any $W_{n}^{(s)}$ window (including Fig. \ref{F2}(botton)),
in which the integer $m^{(s)}$ was introduced to designate the number
of amplifications ($m^{(s)}=j-1$), with $m^{(a)}=0$, $m^{(b)}=1$,
etc.}

\label{F5} 
\end{figure*}

In Fig. \ref{F5} we show details of the collisions for the transmission
windows in Figs. \ref{F4}(b) and (d) and for the reflection windows
in Figs. \ref{F4}(c) and (e). Again, we note the formation of a bound-state
with fixed lifetime for any value of $v_{0}$ within the same window,
and also, collisions of successive windows differs only by one shape-oscillation
period. Thus, the interesting feature observed for the reflection
windows in the first structure extends for any window, where the collision
may be of type transmission or reflection.

The self-similarity involving the structures in Fig. \ref{F4} and
the high sensitivity to the initial conditions, reinforce the hypothesis
that the soliton scattering described by the reduced ODE model is
chaotic and the exit velocity plots are different views of a fractal.
Thus, we found that the reflection and transmission windows are intervals
where the chaotic behavior disappears, that is, for any $v_{0}$ taken
within these windows one can immediately predict the outcome of the
collision, as well as the lifetime of the solitons bound-state.

\subsection{Direct numerical simulations \label{sub:Direct-Numerical-simulations}}

The coupled NLS equations \eqref{cnse1} and \eqref{cnse2} were simulated
by using the split-step method to perform the temporal evolution of
the fields. To solve the linear part of the equations we used the
Crank-Nicholson algorithm while for the nonlinear part we applied
the $4^{\text{th}}$-order Runge-Kutta algorithm. Also, we take the
initial conditions for simulations of solitons' collisions in the
form\begin{subequations}

\begin{align}
\psi(0,T) & =\exp\left\{ \frac{1}{2}\left(1+\frac{1}{g}\right)+g(T-T_{0})^{2}+\frac{i}{4}v_{0}T\right\} ,\label{sol_ic1}\\
\phi(0,T) & =\exp\left\{ \frac{1}{2}\left(1+\frac{1}{g}\right)+g(T+T_{0})^{2}-\frac{i}{4}v_{0}T\right\} ,\label{sol_ic2}
\end{align}
\end{subequations}where the initial separation $2T_{0}$ needs to
be large enough to provide a negligible overlap of the solitons tails
at $T=0$, such that the localized solutions are indeed a good approximation
for the fields at $z=0$. In our case it was taken to be equal to
$20$ dimensionless units. As mentioned before, localized solutions
are obtained for negative values of $g$ in the nonlinear term (in
our case we set $g=-1$). In the collision simulations, the spatial
interval were {[}$-50,50${]}, the $T$-stepsize were $0.04$, and
the $z$-stepsize $0.001$, values that satisfy the CFL condition
associated with the implemented discretization in the linear part
of the coupled NLS equations given by Eqs. \eqref{cnse1} and \eqref{cnse2}.
We verified that for wider spatial intervals there will be no significant
changes for the collision results in the observed bound-state scenarios.
The numeric procedure was implemented with the same numerical precisions
used in the variational approach. In our simulations, we found to
be negligible the losses by radiation due to the approximate solution
profile used as initial conditions. In order to avoid very long simulations,
we analyze the position of both solitons to predict whether the collision
leads to a trapping scenario or not, and if so we assign a null value
to the exit velocity. Several simulations were performed to investigate
the scattering of solitons under the same conditions addressed in
the reduced ODE model. The parameter $v_{0}$ was varied in intervals
with $5000$ grid points, where these intervals were chosen in accordance
with the results obtained in the variational model.

The obtained scattering data shows that three collision scenarios
are possible, \emph{viz.}, transmission, reflection, and trapping
ones. We stress that trapped solitons were not observed in the reduced
ODE model, which was expected because no radiation emission process
is described by that model \cite{Yang_PRL00,Tan_PRE01,Yang_10}. 

\begin{figure}[tb]
\centering \includegraphics[width=1\columnwidth]{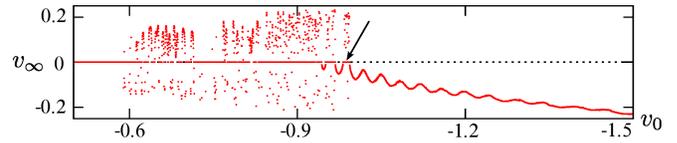} 
\caption{(Color online) Exit-velocity $v_{\infty}$ versus collision-velocity
$v_{0}$ obtained via direct numerical simulations of Eqs. \eqref{cnse1}
and \eqref{cnse2} with initial conditions given by Eqs. \eqref{sol_ic1}
and \eqref{sol_ic2}. The arrow indicates the critical velocity $v_{c}$,
above which (in absolute value) no longer obtains the chaotic pattern.}
\label{F6-1} 
\end{figure}

\begin{figure}[tb]
\centering \includegraphics[width=1\columnwidth]{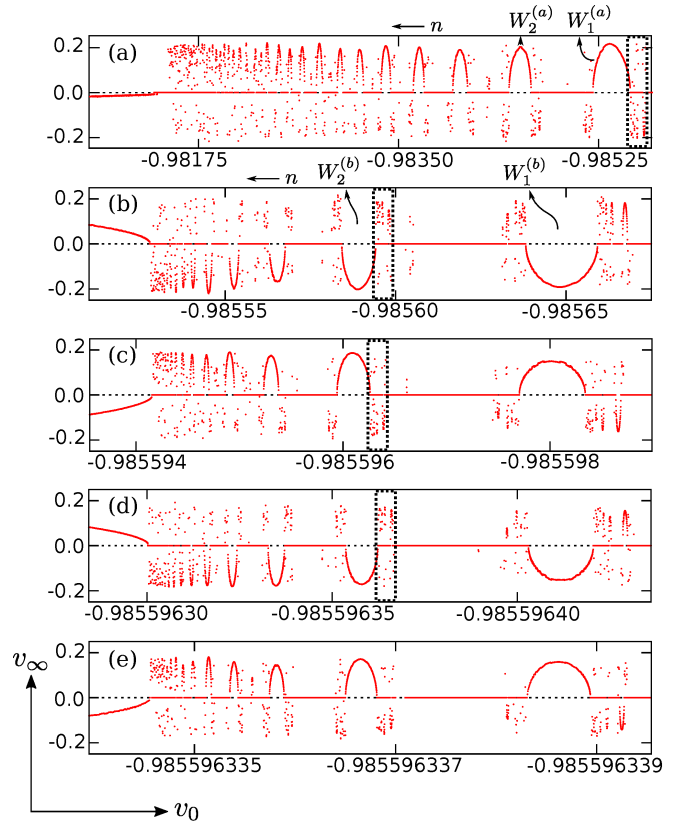} 
\caption{(Color online) Successive amplifications of exit-velocities ($v_{\infty}$)
versus input-velocities ($v_{0}$) due to the scattering of solitons.
(a) Result of the amplification of the tiny interval close to the
arrow (left side) in Fig. \ref{F6-1}; (b)-(e) are successive amplifications
of the intervals highlighted in the exit-velocity graphs (a)-(d),
respectively.}
\label{F6} 
\end{figure}

The first exit velocity graph acquired from the direct numerical simulations
is shown in Fig. \ref{F6-1}, where one can note the existence of
a critical velocity $v_{c}\approx-0.993$ that separates the region
of chaotic scattering from the region of regular scattering. Similarly
to the reduced ODE model, the critical velocity is very close to $-1$
and the collision is elastic only if $v_{0}\ll v_{c}$ ($|v_{0}|\gg|v_{c}|$).
This first view of the exit velocity graph reveals the existence of
reflection and transmission windows with different shapes, where some
of these windows appear distributed uniformly only in a tiny interval
limited by the critical velocity at the right side and the edge of
a transmission window at the left side, which is indicated with an
arrow in Fig. \ref{F6-1} and shown amplified in Fig. \ref{F6}(a).
By analyzing the windows structure revealed by the collisional dynamics
in Fig. \ref{F6}(a), we found the same feature of the reduced ODE
model, that is, the outcome of each collision with initial velocity
within a specific window is always of reflective type, preceded by
a bound-state of same lifetime and number of shape oscillations. Note
that this first amplification yields a windows structure that resembles
the first one provided by the reduced ODE model (Fig. \eqref{F4}(a)).
However, in the numerical simulations we observe a great number of
trapped states, returning null output velocities. Also, we noted a
small difference in the width and position of the windows when comparing
both models. 

\begin{figure*}[tb]
\centering \includegraphics[clip,width=0.4\linewidth]{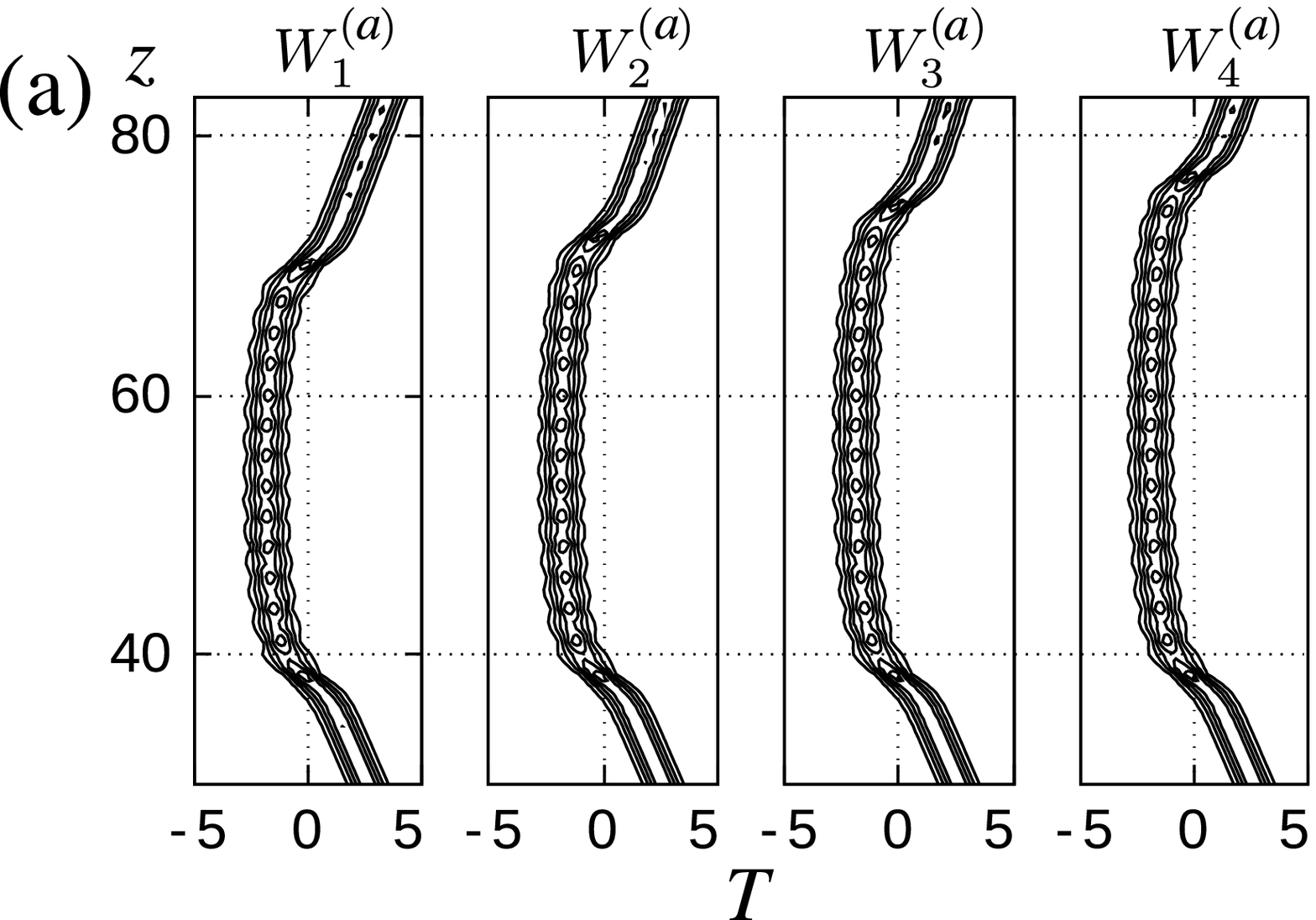}\hfil
\includegraphics[clip,width=0.4\linewidth]{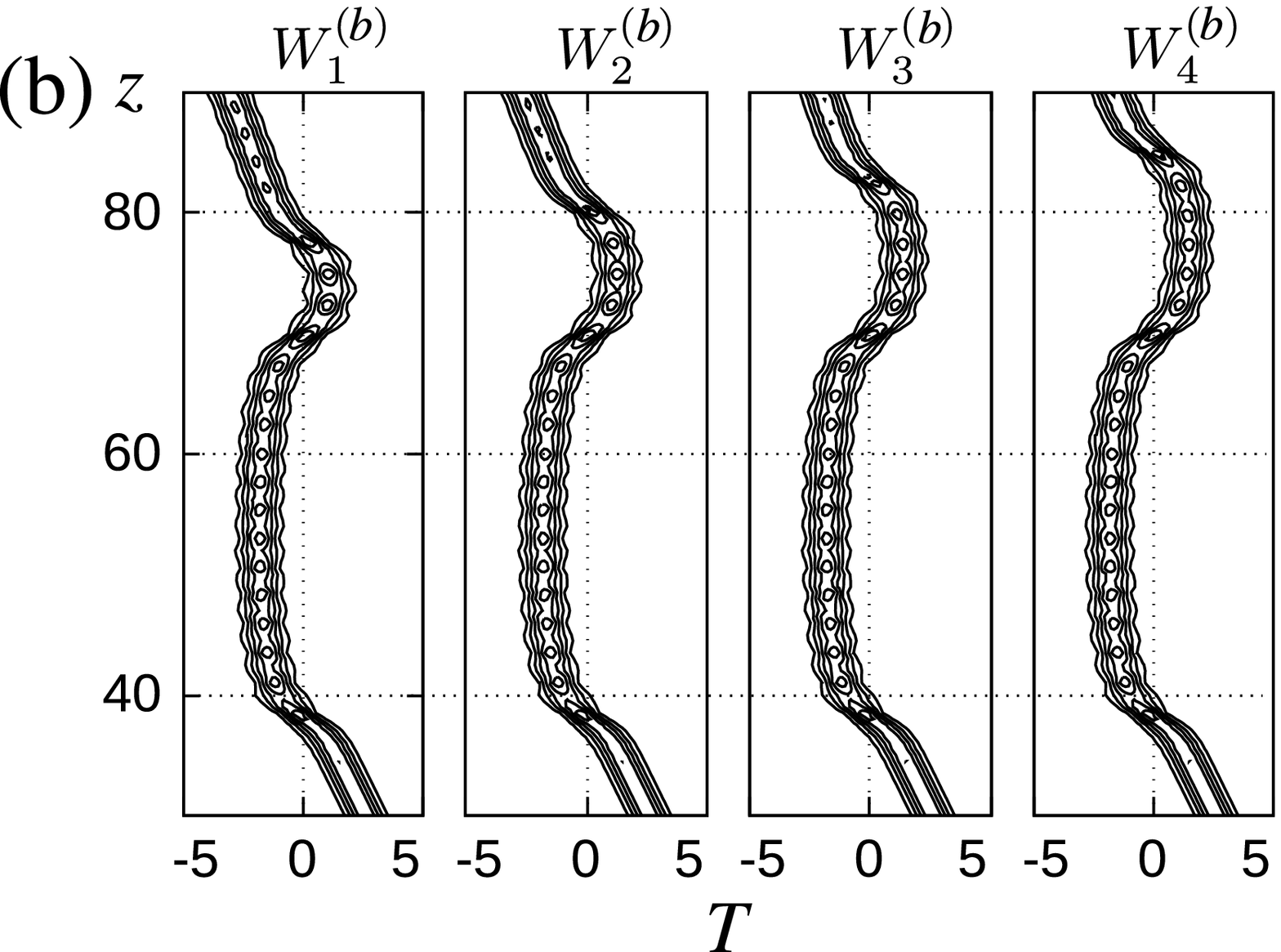}\hfil \includegraphics[clip,width=0.4\linewidth]{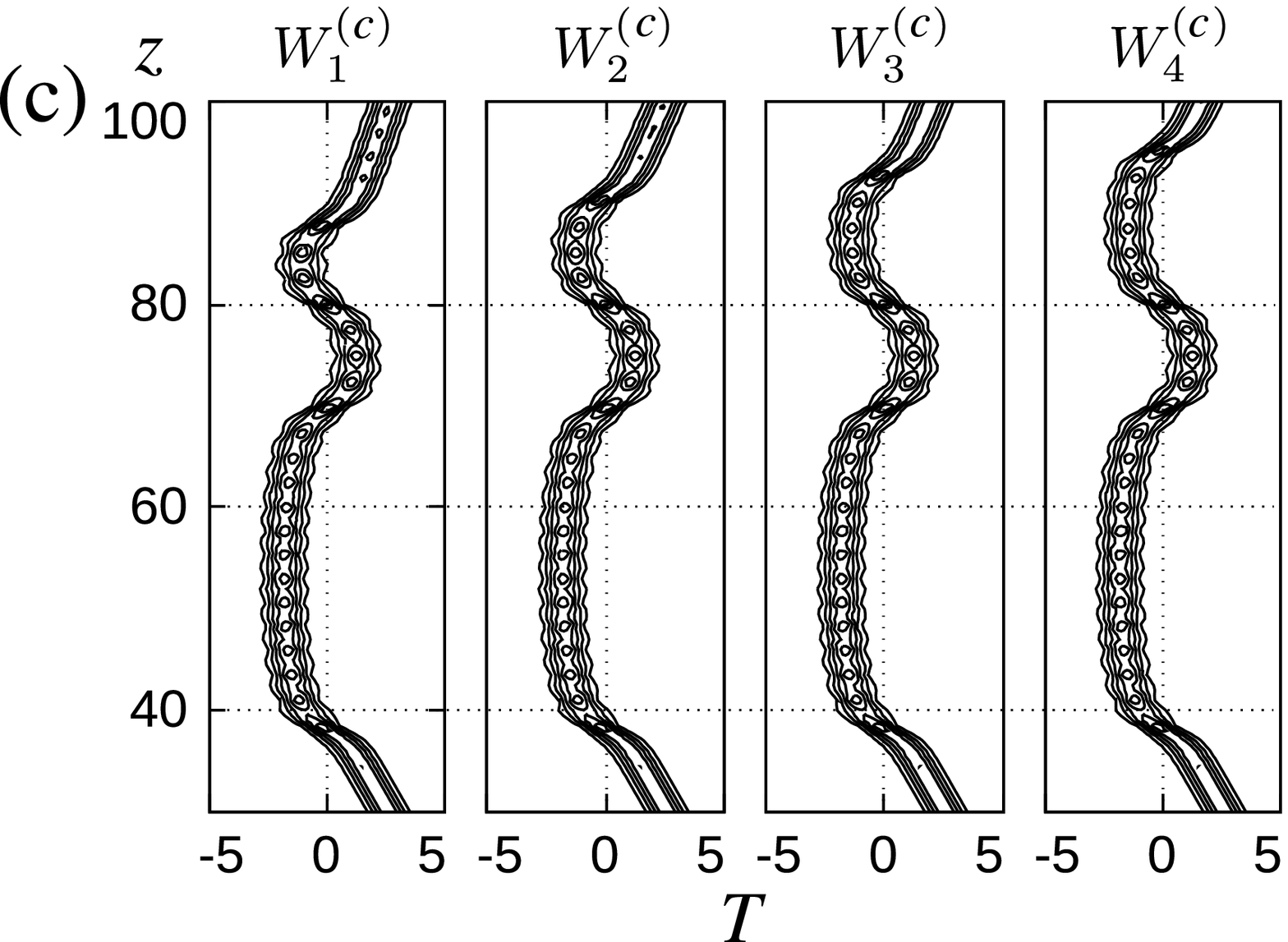}\hfil
\includegraphics[clip,width=0.4\linewidth]{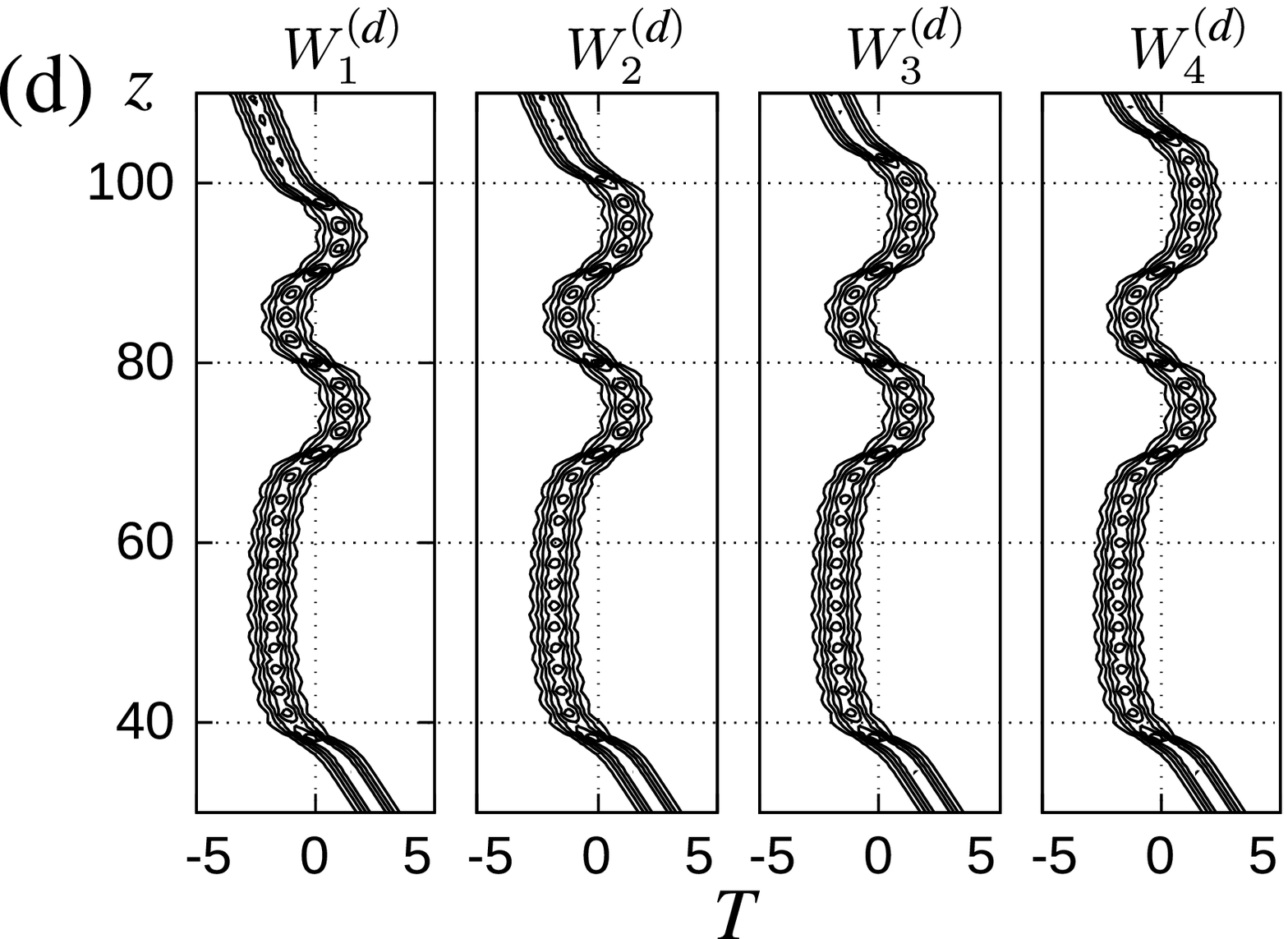}\hfil \includegraphics[clip,width=0.4\linewidth]{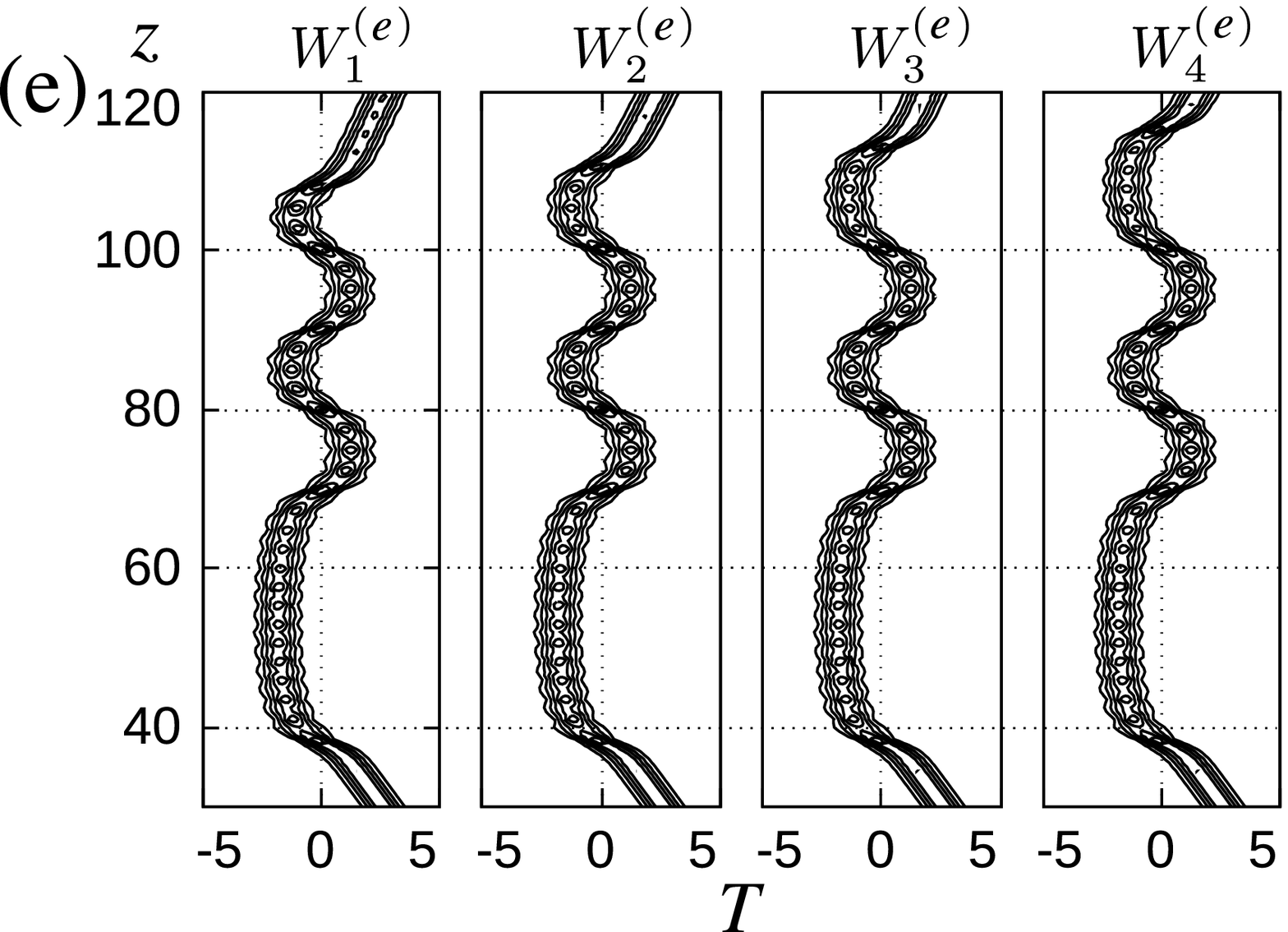} 

\caption{Profile of the localized solution $|\psi(z,T)|^{2}$ during the interaction
with the field $|\phi(z,T)|^{2}$ obtained via direct numerical simulations
of Eqs. \eqref{cnse1} and \eqref{cnse2}. The initial velocity values
are taken within four successive windows in the structure shown in
Fig. \ref{F6}(s), with $s=\{a,b,c,d,e\}$. By counting the number
of shape-oscillations in the solitons bound-state we found $N_{SO}^{n}=(n-1)+m^{(s)}\times4+13+N_{0}$
(where $N_{0}=1\protect\Longleftrightarrow m^{(s)}=0$, otherwise
$N_{0}=0$) for any $W_{n}^{(s)}$ window.}
\label{F7} 
\end{figure*}

The second amplification was taken in the right edge of the largest
window of Fig. \ref{F6}(a), which is highlighted by a dashed rectangle,
corresponding to the plot displayed in Fig. \ref{F6}(b). To construct
the plots of each amplification we performed new simulations for $5000$
different values of input velocities $v_{0}$ within the appropriate
range. Furthermore, Fig. \ref{F6}(b) revealed a structure composed
by transmission windows with different spacing. Indeed, one can note
that this structure is more similar to those obtained by reduced ODE
model than the previous structure. From this point, we proceeded with
amplifications following a protocol equivalent to that employed in
the reduced ODE model simulations. This region allow us to get a symmetry
for each amplification, changing only by a reflection in vertical
axis, which simplifies our analysis of the fractal pattern. Note that
the Fig. \ref{F6}(c) appears approximately as a reflection of Fig.
\ref{F6}(b). We observe this pattern when comparing Figs. \ref{F6}(c)
and \ref{F6}(d), as well as Figs. \ref{F6}(d) and \ref{F6}(e).
In addition, we get very similar patterns revealing a fractal structure
when comparing the amplifications shown in Figs. \ref{F6}(b) and
\ref{F6}(d) and Figs. \ref{F6}(b) and \ref{F6}(d). As we observed
in the reduced ODE model, the bound-state associated with two successive
windows in structures of Figs. \ref{F6}(a)-(e) differs only by one
shape-oscillation, which can be expressed by $N_{SO}^{n+1}=N_{SO}^{n}+1$.
Using the same notation $W_{n}^{(s)}$, we highlighted the first two
windows in the Figs. \ref{F6}(a) and \ref{F6}(b), indicating the
sequence used for the index $n$. In Fig. \ref{F7} we display some
examples of collision scenarios obtained via direct numerical simulations
with values of initial velocities within four successive windows in
each structure shown in Figs. \ref{F6}(a)-(e).

\begin{figure}[tb]
\centering \includegraphics[width=0.85\columnwidth]{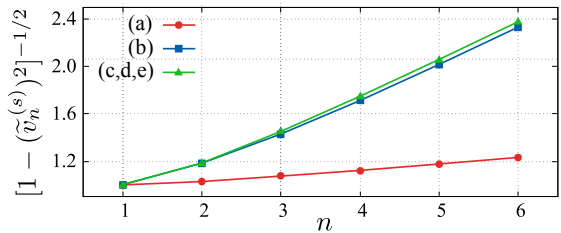}

\caption{(Color online) Plot of $[1-(\widetilde{v}_{n}^{\,(s)})^{2}]^{-1/2}$
as a function of the the window index $n$. In (a)-(e) we used data
obtained by the structures shown in Fig. \ref{F6}(a)-(e), respectively.}
\label{FL2} 
\end{figure}

A similar analysis, previously realized for the window patterns in
the reduced ODE model, was performed here considering the structures
in Fig. \ref{F6}. The relation involving the quantities $[|(v_{n}^{(s)})^{2}-(v_{c}^{(s)})^{2}|]^{-1/2}$
and the window index $n$ is also very well established in accordance
with Eq. (\ref{eq:linear1}), hence the slope $\widetilde{p}^{\,(s)}$
and the intercept coefficient $\widetilde{q}^{\,(s)}$ were obtained
for the structures transformed in the same way described before. We
found $\widetilde{p}^{\,(a)}=0.047\pm0.002\,(4.26\%)$ and $\widetilde{q}^{\,(a)}=0.941\pm0.009\,(0.96\%)$.
Again, for the remaining rescaled structures, we calculated the average
of the coefficients together with their standard deviations, yielding
$\langle\widetilde{p}\rangle=0.271\pm0.006\,(2.23\%)$ and $\langle\widetilde{q}\rangle=0.667\pm0.006\,(0.84\%)$.
Clearly, these values present relevant differences when compared with
$\widetilde{p}^{\,(a)}$ and $\widetilde{q}^{\,(a)}$, respectively.
Therefore, these results provide a numerical evidence that the window
spacing of the structure in Fig. \ref{F6}(a) and the amplified ones
embedded in it are indeed very different when compared. Also, since
$\widetilde{p}^{\,(a)}<\langle\widetilde{p}\rangle$, the window pattern
of the Fig. \ref{F6}(a) contain less spaced windows, as one can see
in the plots. In Fig. \ref{FL2} is displayed the relation between
$[1-(\widetilde{v}_{n}^{\,(s)})^{2}]^{-1/2}$ and the window index
$n$. The remarkably linear relation of these quantities appears very
well established as well as in the result obtained with the reduced
ODE model. Additionally, comparison between the values of $\langle\widetilde{p}\rangle$
yielded by the direct and variational approach shows that they differ
only by $\sim3.2\%$. This reinforces that the reduced ODE model provides
a very good approximation regarding the window pattern. Since the
linear relation in Eq. (\ref{eq:linear1}) it is suitable for the
window spacing description, it can be used as an estimative of the
position of very narrow windows that are difficult to access.

When analyzing the length of the structures given by the direct numerical
simulations, we found that all structures resultant of amplifications
obey with great precision the linear relation given by Eq. \eqref{eq:linear2},
for which the coefficients are $R=-1.481\pm0.006\,(0.43\%)$ and $r=-0.91\pm0.02\,(2.34\%)$.
As shown in Eq. \eqref{eq:length}, only the coefficient $R$ is important
in the zoom ratio. By comparing its value provided by the two approaches
we found a difference of $\sim9.6\%$. However, the first window structure
of the direct numerical simulations has a length that is more than
two orders of magnitude smaller than the length found in the variational
approach, resulting in even smaller structures when amplifying.

\subsection{Analysis of soliton scattering for both approaches \label{sub:Analysis-of-the}}

\begin{figure*}[tb]
\centering \includegraphics[width=0.9\textwidth]{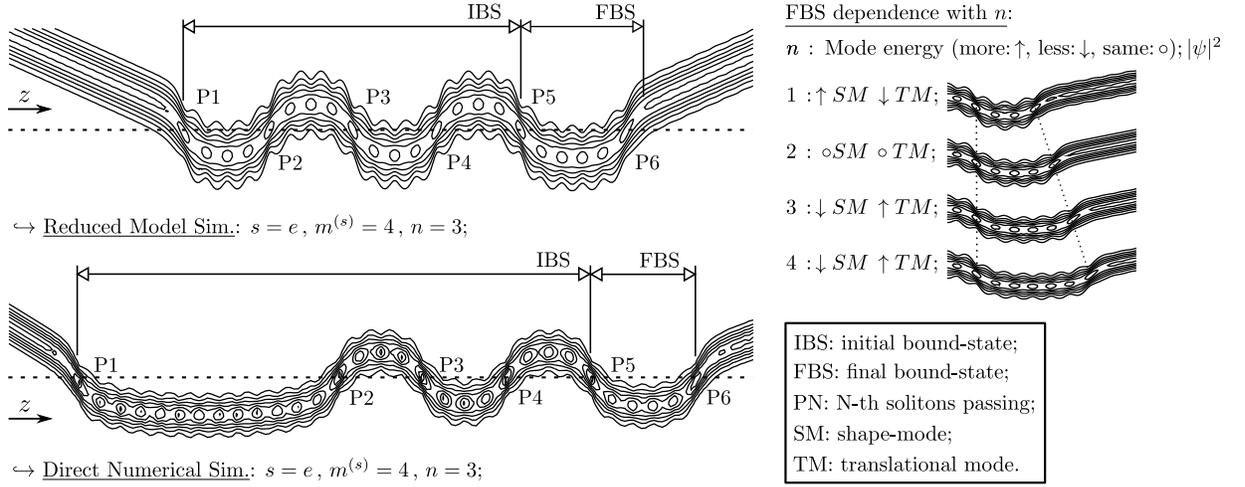}
\caption{Scheme identifying the initial bound state and the final bound-state,
the collisions in the figure are plotted as contour lines of $|\psi(z,T)|^{2}$,
the initial velocity was taken in the $W_{4}^{(e)}$ window for structures
in Fig. \ref{F4}(e) (top) and Fig. \ref{F6}(e) (bottom). In these
collisions the fifth passing (P5) is when the bound-state changes.
At the left side of the figure we highlight how the final bound-state
depends upon $n$, the symbols $\uparrow$, $\downarrow$ and $\circ$
indicate whether energy of the mode has increased, decreased or kept
unchanged, respectively.}
\label{Scheme} 
\end{figure*}

\qquad In order to compare the soliton collision dynamics
given by the two employed methods, we analyze only the interaction
stage that follows after the waves cross the $z$ axis at $T=0$,
where they overlap at most and energy can be exchanged between the
solitons and their internal and external modes. A mechanism of resonance
energy exchange between the internal and translational modes of the
solitons was proposed in Ref. \cite{Campbell_PD83}, where this mechanism
was employed to describe a structure of reflection windows intertwined
by trapping intervals in kink-antikink collisions. The energy exchange
was found to take place at the instants of great overlap of the waves,
which is when an internal mode can become active by storing part of
the kinetic energy. The resonance mechanism was associated to the
existence of fractal window structure resultant in kink-antikink collisions
in the scalar $\lambda(\phi^{2}-1)^{2}$ theory \cite{Anninos_PRD91},
and in Refs. \cite{Yang_10,Yang_PRL00,Tan_PRE01} regarding vector-soliton
collision. Here we also adopt this mechanism to explain how the soliton
exchange kinetic energy with its oscillation-mode, which store energy
in soliton's shape vibrations. In our simulations we noted that this
exchange also occurs when the solitons overlap at most by passing
each other.

For our system, the analysis of the exit velocity plots presented
previously revealed how sensitive can the collision dynamic be. Indeed,
just after the solitons pass each other for the first time, what follows
next depends on the impact velocity that is directly related to $v_{0}$.
Considering the controlled collision scenarios provided by reflection/transmission
windows, in the first passing drained kinetic energy sets up shape-oscillations,
then the attraction interaction binds the solitons that remain in
a bound-state until they bounce (moving forth and back) to pass each
other again. How the energy is exchanged again depends greatly of
the relative phases of the internal modes of the solitons, as discussed
in Ref. \cite{Anninos_PRD91}. If enough kinetic energy is restored
the pair unbinds, otherwise the bound-state prevails. In Figs. \ref{F5}(a)-(d)
and \ref{F7}(b)-(e) the bound-state dynamics are initially the same
for every window in a given structure, considering the bouncing motion
and the number of shape-oscillations, but it changes in a different
manner at a certain passing. In both approaches we verified that these
changes happen when the solitons pass each other $(m^{(s)}+1)$ times
and after that another type of bound-state arises. This final bound-state
precede the unbinding of the solitons and has a varying dynamics,
in which the translational mode can lose energy ($n=1$), keep it
unchanged ($n=2$) or increase it ($n>2$). In any figure regarding
these collisions one can realize what changed in the solitons kinetic
energy by looking at the number of shape-oscillation during a complete
bounce of the solitons. If it was shortened, for example, then more
energy was drained from the translational mode. 

In the direct numerical simulations the initial bound-state has a
longer lifetime and the solitons bounce around the $z$-axis as in
variational approach, but before the second passing time the bound-state
has less energy in the internal modes and more energy in the translational
modes, resulting in a longer bouncing motion as seen in the Figs.
\ref{F7}(b)-(e). The dynamic of this bound-state initially resembles
the collision for the first window in Fig. \ref{F7}(a), which presents
the same number of shape-oscillations. Therefore, the final bound-state
dynamics depends upon $n$, furthermore it changes in the same manner
of the single bound-states formed in the first structure of windows
(in which there is only the final bound-state). This last statement
is valid for both presented approaches. In Fig. \ref{Scheme} we show
a scheme that identify the initial and final bound-state that appear
in a reflectional collision, which is associated to the $W_{4}^{(e)}$
window found after the fourth amplification. We highlight in this
figure how the final bound-state depends upon $n$, and also whether
the internal or translational modes acquire energy after the exchange.

Hitherto we have omitted the losses by radiation, which have an important
role in the solitons' collision dynamics. These losses can be excited
during the instants of energy exchange and carry out energy of the
translation and/or internal modes of the solitons. In Ref. \cite{Anninos_PRD91}
the energy taken by the radiation losses is later retransferred to
the pair, while for vector-soliton collisions they develop a different
role in the dynamics. As mentioned before, although the variational
approach describe very well the resonance mechanism involving the
translational and internal modes, it does not support radiation losses
due to the constraints imposed by the ansatz. This explains the absence
of trapping scenarios in our simulations of the reduced ODE model,
since radiation emission can prevent the translational modes to recover
enough energy to unbind the bound-state of solitons, which does happen
in the direct numerical simulations of the field equations. In these
simulations, all the trapping scenarios are characterized by the solitons
quickly bouncing around the $z$-axis while very close to each other.
In contrast, for a collision described by the reduced ODE model we
attested that after a sufficiently long time eventually occurs a phase
matching of internal modes, retransferring kinetic energy to the solitons,
in a such way that they can immediately unbind or bounce until a final
energy exchange that results in a definitive unbinding.

Since there is no radiation emission in the reduced ODE model, solitons
are able to escape with velocity $|v_{\infty}|\leq|v_{0}|$ in the
variational description, and the height (depth) of the reflection
(transmission) windows is closely equal to the value $|v_{0}|$ corresponding
to the its maximum (minimum). On the other hand, in the direct numerical
simulations radiation emission always happens during the bound-state,
which results in $|v_{\infty}|$ always smaller than $|v_{0}|$ in
all structures. When solitons unbind and move away, the exit velocity
value show us if the internal modes remained excited after the collision
because the smaller the value of $|v_{\infty}|$ is, compared with
$|v_{0}|$, greater is the energy in shape oscillation. Analyzing
the radiation emitted during the collisions in the direct numerical
simulations, we noted that it happens not only when the solitons pass
through each other, but also when they are moving side by side as
a bound-state. This means that such radiation losses play an important
role in the window structure and in the dynamics of the reflectional/transmissional
collisions, and not only in trapping scenarios. 

By a graphical analyses of the solitons' collisions, we found that
its lifetime ($Z_{n}^{(s)}$) is nicely predicted by a multi-linear
function of the integer quantities $n$ and $m^{(s)}$, likewise the
number of shape-oscillations $N_{SO}^{n}$, as shown in Figs. \ref{F5}
and \ref{F7}. They are given by

\begin{eqnarray}
Z_{n}^{(s)} & = & (n-1)z_{f}+m^{(s)}z_{i}+z_{1}\,;\nonumber \\
N_{SO}^{n} & = & (n-1)+4m^{(s)}+4\,;\nonumber \\
 & \hookrightarrow & \begin{cases}
z_{f} & =2.39\pm0.02\,(0.65\%)\\
z_{i} & =10.16\pm0.06\,(0.57\%)\\
z_{1} & =7.9\pm0.1\,(1.26\%),
\end{cases}\label{eq:multi1}
\end{eqnarray}

\noindent for the variational model simulations, and

\begin{eqnarray}
Z_{n}^{(s)} & = & (n-1)z_{f}+m^{(s)}z_{i}+z_{1}\,;\nonumber \\
N_{SO}^{n} & = & (n-1)+4m^{(s)}+13+N_{0}\,,\nonumber \\
 &  & \text{(where \ensuremath{\ensuremath{N_{0}}=\ensuremath{1}\Longleftrightarrow m^{(s)}=0}, otherwise \ensuremath{\ensuremath{N_{0}}=0})\,;}\nonumber \\
 & \hookrightarrow & \begin{cases}
z_{f} & =2.38\pm0.07\,(3.01\%)\\
z_{i} & =9.4\pm0.4\,(4.61\%)\\
z_{1} & =31.5\pm0.9\,(2.79\%),
\end{cases}\label{eq:multi2}
\end{eqnarray}

\noindent for the direct numerical simulations. We point out that
in \eqref{eq:multi2} the relation for $N_{SO}^{n}$ is multi-linear
only for the amplified structures ($m^{(s)}>0$), which is due to
$N_{0}$ that was added to encompass all the structures in a single
formula. In \eqref{eq:multi1} and \eqref{eq:multi2}, $z_{f}$ and
$z_{i}$ are characteristic times of the final bound-state (FBS) and
initial bound-state (IBS) (see Fig. \eqref{Scheme}), respectively,
while $z_{1}$ is the lifetime of the bound-state associated with
the first reflection window of the first window structure (given by
$n=1$ and $m^{(a)}=0$ in both formulas). The coefficient $z_{f}$
represents the width $\Delta z$ of the solitons' shape vibration
in the FBS. The results show that it is closely the same for every
structure, which means that the quantity of energy stored in the internal
mode influences only in the amplitude of the shape vibrations and
not in its frequency. The $z_{i}$ coefficient represents the width
$\Delta z$ of the solitons bounce (shortest bounce in the direct
numerical simulations) in the IBS. The multi-linear relations for
$Z_{n}^{(s)}$ and $N_{SO}^{n}$ are clearly similar, but this is
an expected result since each coefficient in $Z_{n}^{(s)}$ is related
to a number of shape-oscillations. For the reduced ODE model, both
$Z_{n}^{(s)}$ and $N_{SO}^{n}$ can be well predicted for the bound-states
by using the formulas above, while for the direct numerical simulations
they provide a less precise prediction for $Z_{n}^{(s)}$, but still
reasonable. The numerical values of the coefficients in these formulas
give us some useful information about the two approaches. The difference
between the $z_{1}$ coefficients reproduces the difference seen in
the first bounce in the IBS, which resides mainly in the number of
shape-oscillations that is provided by the last coefficient of the
$N_{SO}^{n}$ formulas, where for the variational approach it is $4$
and for the direct numerical simulations it is $14$. The other $z$-coefficients
have similar values. Additionally, the FBS lifetime and shape vibrations
are nicely reproduced by the reduced ODE model.

\section{Conclusion \label{sec:Conclusion}}

In conclusion, we studied the fractal scattering of Gaussian solitons
in directional couplers with logarithmic nonlinearities. In this sense,
we employed two methods, viz., the variational approach and direct
numerical simulations. Regarding the variational approach, we have
started our study by firstly developing the reduced ODE model for
our system governed by the field equations \eqref{cnse1} and \eqref{cnse2},
which provided the results presented in the section \ref{sec:Theoretical-Model}.
The reduced ODE model description shows that the collisions of the
solitons is chaotic when the absolute value of the input velocity
is less than a certain critical value, and a fractal structure composed
by reflection and transmission windows arises within the chaotic region.
In the same way, in view to verify the feasibility of the reduced
ODE model in the present context we performed direct numerical simulations.
So, in Sec. \ref{sec:Numerical-Results} we shown similar features
presented by the two approaches employed. 

Our analysis on the size of the structures and window positions yielded
quite precise results for the fitting curves. As long as the amplifications
follow the adopted protocol these structures closely preserve their
window pattern and they are all embedded in each other in a quite
well defined manner, which means that the amplifications almost occur
in the same ratio. Interestingly, a numerical comparison of the window
pattern and amplification ratio of both variational and direct numerical
approaches shows that these are not very much different. Once again,
it reinforces the usefulness of the variational model in predicting
the major features of solitons scattering and collision dynamics.

This study also gives us an idea about the form of collisions of solitons
in reflection/transmission windows within the chaotic region, which
can enable us to control the pattern of collision via initial approach
velocity between them. Also, due to sensitivity to small changes in
the velocity, we theorize that it could be used as a kind of sensor
to verify inhomogeneities caused by impurities in the medium. In this
sense, we are studying other models as well as including tests of
effects of inhomogeneities in the medium.
\begin{acknowledgments}
We acknowledge financial support from the Brazilian agencies CNPq,
CAPES, and the National Institute of Science and Technology (INCT)
for Quantum Information.
\end{acknowledgments}

\bibliographystyle{apsrev4-1}
\bibliography{Refs}

\begin{thebibliography}{58}%
\makeatletter
\providecommand \@ifxundefined [1]{%
 \@ifx{#1\undefined}
}%
\providecommand \@ifnum [1]{%
 \ifnum #1\expandafter \@firstoftwo
 \else \expandafter \@secondoftwo
 \fi
}%
\providecommand \@ifx [1]{%
 \ifx #1\expandafter \@firstoftwo
 \else \expandafter \@secondoftwo
 \fi
}%
\providecommand \natexlab [1]{#1}%
\providecommand \enquote  [1]{``#1''}%
\providecommand \bibnamefont  [1]{#1}%
\providecommand \bibfnamefont [1]{#1}%
\providecommand \citenamefont [1]{#1}%
\providecommand \href@noop [0]{\@secondoftwo}%
\providecommand \href [0]{\begingroup \@sanitize@url \@href}%
\providecommand \@href[1]{\@@startlink{#1}\@@href}%
\providecommand \@@href[1]{\endgroup#1\@@endlink}%
\providecommand \@sanitize@url [0]{\catcode `\\12\catcode `\$12\catcode
  `\&12\catcode `\#12\catcode `\^12\catcode `\_12\catcode `\%12\relax}%
\providecommand \@@startlink[1]{}%
\providecommand \@@endlink[0]{}%
\providecommand \url  [0]{\begingroup\@sanitize@url \@url }%
\providecommand \@url [1]{\endgroup\@href {#1}{\urlprefix }}%
\providecommand \urlprefix  [0]{URL }%
\providecommand \Eprint [0]{\href }%
\providecommand \doibase [0]{http://dx.doi.org/}%
\providecommand \selectlanguage [0]{\@gobble}%
\providecommand \bibinfo  [0]{\@secondoftwo}%
\providecommand \bibfield  [0]{\@secondoftwo}%
\providecommand \translation [1]{[#1]}%
\providecommand \BibitemOpen [0]{}%
\providecommand \bibitemStop [0]{}%
\providecommand \bibitemNoStop [0]{.\EOS\space}%
\providecommand \EOS [0]{\spacefactor3000\relax}%
\providecommand \BibitemShut  [1]{\csname bibitem#1\endcsname}%
\let\auto@bib@innerbib\@empty
%</preamble>
\bibitem [{\citenamefont {Zabusky}\ and\ \citenamefont
  {Kruskal}(1965)}]{Zabusky_PRL65}%
  \BibitemOpen
  \bibfield  {author} {\bibinfo {author} {\bibfnamefont {N.~J.}\ \bibnamefont
  {Zabusky}}\ and\ \bibinfo {author} {\bibfnamefont {M.~D.}\ \bibnamefont
  {Kruskal}},\ }\href {\doibase 10.1103/PhysRevLett.15.240} {\bibfield
  {journal} {\bibinfo  {journal} {Phys. Rev. Lett.}\ }\textbf {\bibinfo
  {volume} {15}},\ \bibinfo {pages} {240} (\bibinfo {year} {1965})}\BibitemShut
  {NoStop}%
\bibitem [{\citenamefont {Khaykovich}(2002)}]{Khaykovich_SCI02}%
  \BibitemOpen
  \bibfield  {author} {\bibinfo {author} {\bibfnamefont {L.}~\bibnamefont
  {Khaykovich}},\ }\href {\doibase 10.1126/science.1071021} {\bibfield
  {journal} {\bibinfo  {journal} {Science (80-. ).}\ }\textbf {\bibinfo
  {volume} {296}},\ \bibinfo {pages} {1290} (\bibinfo {year}
  {2002})}\BibitemShut {NoStop}%
\bibitem [{\citenamefont {Strecker}\ \emph {et~al.}(2002)\citenamefont
  {Strecker}, \citenamefont {Partridge}, \citenamefont {Truscott},\ and\
  \citenamefont {Hulet}}]{Strecker_NAT02}%
  \BibitemOpen
  \bibfield  {author} {\bibinfo {author} {\bibfnamefont {K.~E.}\ \bibnamefont
  {Strecker}}, \bibinfo {author} {\bibfnamefont {G.~B.}\ \bibnamefont
  {Partridge}}, \bibinfo {author} {\bibfnamefont {A.~G.}\ \bibnamefont
  {Truscott}}, \ and\ \bibinfo {author} {\bibfnamefont {R.~G.}\ \bibnamefont
  {Hulet}},\ }\href {\doibase 10.1038/nature747} {\bibfield  {journal}
  {\bibinfo  {journal} {Nature}\ }\textbf {\bibinfo {volume} {417}},\ \bibinfo
  {pages} {150} (\bibinfo {year} {2002})}\BibitemShut {NoStop}%
\bibitem [{\citenamefont {Cornish}\ \emph {et~al.}(2006)\citenamefont
  {Cornish}, \citenamefont {Thompson},\ and\ \citenamefont
  {Wieman}}]{Cornish_PRL06}%
  \BibitemOpen
  \bibfield  {author} {\bibinfo {author} {\bibfnamefont {S.~L.}\ \bibnamefont
  {Cornish}}, \bibinfo {author} {\bibfnamefont {S.~T.}\ \bibnamefont
  {Thompson}}, \ and\ \bibinfo {author} {\bibfnamefont {C.~E.}\ \bibnamefont
  {Wieman}},\ }\href {\doibase 10.1103/PhysRevLett.96.170401} {\bibfield
  {journal} {\bibinfo  {journal} {Phys. Rev. Lett.}\ }\textbf {\bibinfo
  {volume} {96}},\ \bibinfo {pages} {170401} (\bibinfo {year}
  {2006})}\BibitemShut {NoStop}%
\bibitem [{\citenamefont {Marchant}\ \emph {et~al.}(2013)\citenamefont
  {Marchant}, \citenamefont {Billam}, \citenamefont {Wiles}, \citenamefont
  {Yu}, \citenamefont {Gardiner},\ and\ \citenamefont
  {Cornish}}]{Marchant_NC13}%
  \BibitemOpen
  \bibfield  {author} {\bibinfo {author} {\bibfnamefont {A.~L.}\ \bibnamefont
  {Marchant}}, \bibinfo {author} {\bibfnamefont {T.~P.}\ \bibnamefont
  {Billam}}, \bibinfo {author} {\bibfnamefont {T.~P.}\ \bibnamefont {Wiles}},
  \bibinfo {author} {\bibfnamefont {M.~M.~H.}\ \bibnamefont {Yu}}, \bibinfo
  {author} {\bibfnamefont {S.~A.}\ \bibnamefont {Gardiner}}, \ and\ \bibinfo
  {author} {\bibfnamefont {S.~L.}\ \bibnamefont {Cornish}},\ }\href {\doibase
  10.1038/ncomms2893} {\bibfield  {journal} {\bibinfo  {journal} {Nat.
  Commun.}\ }\textbf {\bibinfo {volume} {4}},\ \bibinfo {pages} {1865}
  (\bibinfo {year} {2013})}\BibitemShut {NoStop}%
\bibitem [{\citenamefont {Burger}\ \emph {et~al.}(1999)\citenamefont {Burger},
  \citenamefont {Bongs}, \citenamefont {Dettmer}, \citenamefont {Ertmer},
  \citenamefont {Sengstock}, \citenamefont {Sanpera}, \citenamefont
  {Shlyapnikov},\ and\ \citenamefont {Lewenstein}}]{Burger_PRL99}%
  \BibitemOpen
  \bibfield  {author} {\bibinfo {author} {\bibfnamefont {S.}~\bibnamefont
  {Burger}}, \bibinfo {author} {\bibfnamefont {K.}~\bibnamefont {Bongs}},
  \bibinfo {author} {\bibfnamefont {S.}~\bibnamefont {Dettmer}}, \bibinfo
  {author} {\bibfnamefont {W.}~\bibnamefont {Ertmer}}, \bibinfo {author}
  {\bibfnamefont {K.}~\bibnamefont {Sengstock}}, \bibinfo {author}
  {\bibfnamefont {A.}~\bibnamefont {Sanpera}}, \bibinfo {author} {\bibfnamefont
  {G.~V.}\ \bibnamefont {Shlyapnikov}}, \ and\ \bibinfo {author} {\bibfnamefont
  {M.}~\bibnamefont {Lewenstein}},\ }\href {\doibase
  10.1103/PhysRevLett.83.5198} {\bibfield  {journal} {\bibinfo  {journal}
  {Phys. Rev. Lett.}\ }\textbf {\bibinfo {volume} {83}},\ \bibinfo {pages}
  {5198} (\bibinfo {year} {1999})}\BibitemShut {NoStop}%
\bibitem [{\citenamefont {Craig}\ \emph {et~al.}(2006)\citenamefont {Craig},
  \citenamefont {Guyenne}, \citenamefont {Hammack}, \citenamefont {Henderson},\
  and\ \citenamefont {Sulem}}]{Craig_PF06}%
  \BibitemOpen
  \bibfield  {author} {\bibinfo {author} {\bibfnamefont {W.}~\bibnamefont
  {Craig}}, \bibinfo {author} {\bibfnamefont {P.}~\bibnamefont {Guyenne}},
  \bibinfo {author} {\bibfnamefont {J.}~\bibnamefont {Hammack}}, \bibinfo
  {author} {\bibfnamefont {D.}~\bibnamefont {Henderson}}, \ and\ \bibinfo
  {author} {\bibfnamefont {C.}~\bibnamefont {Sulem}},\ }\href {\doibase
  10.1063/1.2205916} {\bibfield  {journal} {\bibinfo  {journal} {Phys. Fluids}\
  }\textbf {\bibinfo {volume} {18}},\ \bibinfo {pages} {057106} (\bibinfo
  {year} {2006})}\BibitemShut {NoStop}%
\bibitem [{\citenamefont {Davydov}(1985)}]{Davydov_85}%
  \BibitemOpen
  \bibfield  {author} {\bibinfo {author} {\bibfnamefont {A.~S.}\ \bibnamefont
  {Davydov}},\ }\href {https://books.google.com.br/books?id=8Yy0AAAAIAAJ}
  {\emph {\bibinfo {title} {{Solitons in Molecular Systems}}}},\ Mathematics
  and its applications (D. Reidel Publishing Company).: Soviet series\
  (\bibinfo  {publisher} {D. Reidel Publishing Company},\ \bibinfo {year}
  {1985})\BibitemShut {NoStop}%
\bibitem [{\citenamefont {Yakushevich}(2004)}]{Yakushevich_04}%
  \BibitemOpen
  \bibfield  {author} {\bibinfo {author} {\bibfnamefont {L.~V.}\ \bibnamefont
  {Yakushevich}},\ }\href {https://books.google.com.br/books?id=AHrRLSZYok4C}
  {\emph {\bibinfo {title} {{Nonlinear Physics of DNA}}}}\ (\bibinfo
  {publisher} {Wiley},\ \bibinfo {year} {2004})\BibitemShut {NoStop}%
\bibitem [{\citenamefont {Agrawal}(2001)}]{Agrawal_01}%
  \BibitemOpen
  \bibfield  {author} {\bibinfo {author} {\bibfnamefont {G.}~\bibnamefont
  {Agrawal}},\ }\href {https://books.google.com.br/books?id=wjHP0oAVcScC}
  {\emph {\bibinfo {title} {{Nonlinear Fiber Optics}}}},\ Optics and Photonics\
  (\bibinfo  {publisher} {Elsevier Science},\ \bibinfo {year}
  {2001})\BibitemShut {NoStop}%
\bibitem [{\citenamefont {Hasegawa}\ and\ \citenamefont
  {Kodama}(1995)}]{Hasegawa_95}%
  \BibitemOpen
  \bibfield  {author} {\bibinfo {author} {\bibfnamefont {A.}~\bibnamefont
  {Hasegawa}}\ and\ \bibinfo {author} {\bibfnamefont {Y.}~\bibnamefont
  {Kodama}},\ }\href {https://books.google.com.br/books?id=Y{\_}9SAAAAMAAJ}
  {\emph {\bibinfo {title} {{Solitons in optical communications}}}},\ Oxford
  series in optical and imaging sciences\ (\bibinfo  {publisher} {Clarendon
  Press},\ \bibinfo {year} {1995})\BibitemShut {NoStop}%
\bibitem [{\citenamefont {Bjorkholm}\ and\ \citenamefont
  {Ashkin}(1974)}]{Bjorkholm_PRL74}%
  \BibitemOpen
  \bibfield  {author} {\bibinfo {author} {\bibfnamefont {J.~E.}\ \bibnamefont
  {Bjorkholm}}\ and\ \bibinfo {author} {\bibfnamefont {A.~A.}\ \bibnamefont
  {Ashkin}},\ }\href {\doibase 10.1103/PhysRevLett.32.129} {\bibfield
  {journal} {\bibinfo  {journal} {Phys. Rev. Lett.}\ }\textbf {\bibinfo
  {volume} {32}},\ \bibinfo {pages} {129} (\bibinfo {year} {1974})}\BibitemShut
  {NoStop}%
\bibitem [{\citenamefont {Barthelemy}\ \emph {et~al.}(1985)\citenamefont
  {Barthelemy}, \citenamefont {Maneuf},\ and\ \citenamefont
  {Froehly}}]{Barthelemy_OC85}%
  \BibitemOpen
  \bibfield  {author} {\bibinfo {author} {\bibfnamefont {A.}~\bibnamefont
  {Barthelemy}}, \bibinfo {author} {\bibfnamefont {S.}~\bibnamefont {Maneuf}},
  \ and\ \bibinfo {author} {\bibfnamefont {C.}~\bibnamefont {Froehly}},\ }\href
  {\doibase 10.1016/0030-4018(85)90047-1} {\bibfield  {journal} {\bibinfo
  {journal} {Opt. Commun.}\ }\textbf {\bibinfo {volume} {55}},\ \bibinfo
  {pages} {201} (\bibinfo {year} {1985})}\BibitemShut {NoStop}%
\bibitem [{\citenamefont {Segev}\ \emph {et~al.}(1992)\citenamefont {Segev},
  \citenamefont {Crosignani}, \citenamefont {Yariv},\ and\ \citenamefont
  {Fischer}}]{Segev_PRL92}%
  \BibitemOpen
  \bibfield  {author} {\bibinfo {author} {\bibfnamefont {M.}~\bibnamefont
  {Segev}}, \bibinfo {author} {\bibfnamefont {B.}~\bibnamefont {Crosignani}},
  \bibinfo {author} {\bibfnamefont {A.}~\bibnamefont {Yariv}}, \ and\ \bibinfo
  {author} {\bibfnamefont {B.}~\bibnamefont {Fischer}},\ }\href {\doibase
  10.1103/PhysRevLett.68.923} {\bibfield  {journal} {\bibinfo  {journal} {Phys.
  Rev. Lett.}\ }\textbf {\bibinfo {volume} {68}},\ \bibinfo {pages} {923}
  (\bibinfo {year} {1992})}\BibitemShut {NoStop}%
\bibitem [{\citenamefont {Aitchison}\ \emph {et~al.}(1992)\citenamefont
  {Aitchison}, \citenamefont {Al-Hemyari}, \citenamefont {Ironside},
  \citenamefont {Grant},\ and\ \citenamefont {Sibbett}}]{Aitchison_EL92}%
  \BibitemOpen
  \bibfield  {author} {\bibinfo {author} {\bibfnamefont {J.}~\bibnamefont
  {Aitchison}}, \bibinfo {author} {\bibfnamefont {K.}~\bibnamefont
  {Al-Hemyari}}, \bibinfo {author} {\bibfnamefont {C.}~\bibnamefont
  {Ironside}}, \bibinfo {author} {\bibfnamefont {R.}~\bibnamefont {Grant}}, \
  and\ \bibinfo {author} {\bibfnamefont {W.}~\bibnamefont {Sibbett}},\ }\href
  {\doibase 10.1049/el:19921203} {\bibfield  {journal} {\bibinfo  {journal}
  {Electron. Lett.}\ }\textbf {\bibinfo {volume} {28}},\ \bibinfo {pages}
  {1879} (\bibinfo {year} {1992})}\BibitemShut {NoStop}%
\bibitem [{\citenamefont {Beeckman}\ \emph {et~al.}(2004)\citenamefont
  {Beeckman}, \citenamefont {Neyts}, \citenamefont {Hutsebaut}, \citenamefont
  {Cambournac},\ and\ \citenamefont {Haelterman}}]{Beeckman_OE04}%
  \BibitemOpen
  \bibfield  {author} {\bibinfo {author} {\bibfnamefont {J.}~\bibnamefont
  {Beeckman}}, \bibinfo {author} {\bibfnamefont {K.}~\bibnamefont {Neyts}},
  \bibinfo {author} {\bibfnamefont {X.}~\bibnamefont {Hutsebaut}}, \bibinfo
  {author} {\bibfnamefont {C.}~\bibnamefont {Cambournac}}, \ and\ \bibinfo
  {author} {\bibfnamefont {M.}~\bibnamefont {Haelterman}},\ }\href {\doibase
  10.1364/OPEX.12.001011} {\bibfield  {journal} {\bibinfo  {journal} {Opt.
  Express}\ }\textbf {\bibinfo {volume} {12}},\ \bibinfo {pages} {1011}
  (\bibinfo {year} {2004})}\BibitemShut {NoStop}%
\bibitem [{\citenamefont {Kivshar}\ and\ \citenamefont
  {Agrawal}(2003)}]{Kivshar_03}%
  \BibitemOpen
  \bibfield  {author} {\bibinfo {author} {\bibfnamefont {Y.~S.}\ \bibnamefont
  {Kivshar}}\ and\ \bibinfo {author} {\bibfnamefont {G.}~\bibnamefont
  {Agrawal}},\ }\href {https://books.google.com.br/books?id=zzWgibj4ypsC}
  {\emph {\bibinfo {title} {{Optical Solitons: From Fibers to Photonic
  Crystals}}}}\ (\bibinfo  {publisher} {Elsevier Science},\ \bibinfo {year}
  {2003})\BibitemShut {NoStop}%
\bibitem [{\citenamefont {Zakharov}\ and\ \citenamefont
  {Shabat}(1972)}]{Zakharov_JETP72}%
  \BibitemOpen
  \bibfield  {author} {\bibinfo {author} {\bibfnamefont {V.}~\bibnamefont
  {Zakharov}}\ and\ \bibinfo {author} {\bibfnamefont {A.}~\bibnamefont
  {Shabat}},\ }\href {http://www.jetp.ac.ru/cgi-bin/e/index/e/34/1/p62?a=list}
  {\bibfield  {journal} {\bibinfo  {journal} {Sov. J. Exp. Theor. Phys.}\
  }\textbf {\bibinfo {volume} {34}},\ \bibinfo {pages} {62} (\bibinfo {year}
  {1972})}\BibitemShut {NoStop}%
\bibitem [{\citenamefont {Yang}\ and\ \citenamefont {Tan}(2000)}]{Yang_PRL00}%
  \BibitemOpen
  \bibfield  {author} {\bibinfo {author} {\bibfnamefont {J.}~\bibnamefont
  {Yang}}\ and\ \bibinfo {author} {\bibfnamefont {Y.}~\bibnamefont {Tan}},\
  }\href {\doibase 10.1103/PhysRevLett.85.3624} {\bibfield  {journal} {\bibinfo
   {journal} {Phys. Rev. Lett.}\ }\textbf {\bibinfo {volume} {85}},\ \bibinfo
  {pages} {3624} (\bibinfo {year} {2000})}\BibitemShut {NoStop}%
\bibitem [{\citenamefont {Tan}\ and\ \citenamefont {Yang}(2001)}]{Tan_PRE01}%
  \BibitemOpen
  \bibfield  {author} {\bibinfo {author} {\bibfnamefont {Y.}~\bibnamefont
  {Tan}}\ and\ \bibinfo {author} {\bibfnamefont {J.}~\bibnamefont {Yang}},\
  }\href {\doibase 10.1103/PhysRevE.64.056616} {\bibfield  {journal} {\bibinfo
  {journal} {Phys. Rev. E}\ }\textbf {\bibinfo {volume} {64}},\ \bibinfo
  {pages} {056616} (\bibinfo {year} {2001})}\BibitemShut {NoStop}%
\bibitem [{\citenamefont {Dmitriev}\ and\ \citenamefont
  {Shigenari}(2002)}]{Dmitriev_CHAOS02}%
  \BibitemOpen
  \bibfield  {author} {\bibinfo {author} {\bibfnamefont {S.~V.}\ \bibnamefont
  {Dmitriev}}\ and\ \bibinfo {author} {\bibfnamefont {T.}~\bibnamefont
  {Shigenari}},\ }\href {\doibase 10.1063/1.1476951} {\bibfield  {journal}
  {\bibinfo  {journal} {Chaos An Interdiscip. J. Nonlinear Sci.}\ }\textbf
  {\bibinfo {volume} {12}},\ \bibinfo {pages} {324} (\bibinfo {year}
  {2002})}\BibitemShut {NoStop}%
\bibitem [{\citenamefont {Zhu}\ and\ \citenamefont {Yang}(2007)}]{Zhu_PRE07}%
  \BibitemOpen
  \bibfield  {author} {\bibinfo {author} {\bibfnamefont {Y.}~\bibnamefont
  {Zhu}}\ and\ \bibinfo {author} {\bibfnamefont {J.}~\bibnamefont {Yang}},\
  }\href {\doibase 10.1103/PhysRevE.75.036605} {\bibfield  {journal} {\bibinfo
  {journal} {Phys. Rev. E}\ }\textbf {\bibinfo {volume} {75}},\ \bibinfo
  {pages} {036605} (\bibinfo {year} {2007})}\BibitemShut {NoStop}%
\bibitem [{\citenamefont {Zhu}\ \emph {et~al.}(2008{\natexlab{a}})\citenamefont
  {Zhu}, \citenamefont {Haberman},\ and\ \citenamefont {Yang}}]{Zhu_PRL08}%
  \BibitemOpen
  \bibfield  {author} {\bibinfo {author} {\bibfnamefont {Y.}~\bibnamefont
  {Zhu}}, \bibinfo {author} {\bibfnamefont {R.}~\bibnamefont {Haberman}}, \
  and\ \bibinfo {author} {\bibfnamefont {J.}~\bibnamefont {Yang}},\ }\href
  {\doibase 10.1103/PhysRevLett.100.143901} {\bibfield  {journal} {\bibinfo
  {journal} {Phys. Rev. Lett.}\ }\textbf {\bibinfo {volume} {100}},\ \bibinfo
  {pages} {143901} (\bibinfo {year} {2008}{\natexlab{a}})}\BibitemShut
  {NoStop}%
\bibitem [{\citenamefont {Zhu}\ \emph {et~al.}(2008{\natexlab{b}})\citenamefont
  {Zhu}, \citenamefont {Haberman},\ and\ \citenamefont {Yang}}]{Zhu_PD08}%
  \BibitemOpen
  \bibfield  {author} {\bibinfo {author} {\bibfnamefont {Y.}~\bibnamefont
  {Zhu}}, \bibinfo {author} {\bibfnamefont {R.}~\bibnamefont {Haberman}}, \
  and\ \bibinfo {author} {\bibfnamefont {J.}~\bibnamefont {Yang}},\ }\href
  {\doibase 10.1016/j.physd.2008.03.030} {\bibfield  {journal} {\bibinfo
  {journal} {Phys. D Nonlinear Phenom.}\ }\textbf {\bibinfo {volume} {237}},\
  \bibinfo {pages} {2411} (\bibinfo {year} {2008}{\natexlab{b}})}\BibitemShut
  {NoStop}%
\bibitem [{\citenamefont {Zhu}\ \emph {et~al.}(2009)\citenamefont {Zhu},
  \citenamefont {Haberman},\ and\ \citenamefont {Yang}}]{Zhu_SAM09}%
  \BibitemOpen
  \bibfield  {author} {\bibinfo {author} {\bibfnamefont {Y.}~\bibnamefont
  {Zhu}}, \bibinfo {author} {\bibfnamefont {R.}~\bibnamefont {Haberman}}, \
  and\ \bibinfo {author} {\bibfnamefont {J.}~\bibnamefont {Yang}},\ }\href
  {\doibase 10.1111/j.1467-9590.2009.00442.x} {\bibfield  {journal} {\bibinfo
  {journal} {Stud. Appl. Math.}\ }\textbf {\bibinfo {volume} {122}},\ \bibinfo
  {pages} {449} (\bibinfo {year} {2009})}\BibitemShut {NoStop}%
\bibitem [{\citenamefont {Hause}\ \emph {et~al.}(2010)\citenamefont {Hause},
  \citenamefont {Hartwig},\ and\ \citenamefont {Mitschke}}]{Hause_PRA10}%
  \BibitemOpen
  \bibfield  {author} {\bibinfo {author} {\bibfnamefont {A.}~\bibnamefont
  {Hause}}, \bibinfo {author} {\bibfnamefont {H.}~\bibnamefont {Hartwig}}, \
  and\ \bibinfo {author} {\bibfnamefont {F.}~\bibnamefont {Mitschke}},\ }\href
  {\doibase 10.1103/PhysRevA.82.053833} {\bibfield  {journal} {\bibinfo
  {journal} {Phys. Rev. A}\ }\textbf {\bibinfo {volume} {82}},\ \bibinfo
  {pages} {053833} (\bibinfo {year} {2010})}\BibitemShut {NoStop}%
\bibitem [{\citenamefont {Goodman}(2008)}]{Goodman_CHAOS08}%
  \BibitemOpen
  \bibfield  {author} {\bibinfo {author} {\bibfnamefont {R.~H.}\ \bibnamefont
  {Goodman}},\ }\href {\doibase 10.1063/1.2904823} {\bibfield  {journal}
  {\bibinfo  {journal} {Chaos An Interdiscip. J. Nonlinear Sci.}\ }\textbf
  {\bibinfo {volume} {18}},\ \bibinfo {pages} {023113} (\bibinfo {year}
  {2008})}\BibitemShut {NoStop}%
\bibitem [{\citenamefont {Goodman}\ \emph {et~al.}(2015)\citenamefont
  {Goodman}, \citenamefont {Rahman}, \citenamefont {Bellanich},\ and\
  \citenamefont {Morrison}}]{Goodman_CHAOS15}%
  \BibitemOpen
  \bibfield  {author} {\bibinfo {author} {\bibfnamefont {R.~H.}\ \bibnamefont
  {Goodman}}, \bibinfo {author} {\bibfnamefont {A.}~\bibnamefont {Rahman}},
  \bibinfo {author} {\bibfnamefont {M.~J.}\ \bibnamefont {Bellanich}}, \ and\
  \bibinfo {author} {\bibfnamefont {C.~N.}\ \bibnamefont {Morrison}},\ }\href
  {\doibase 10.1063/1.4917047} {\bibfield  {journal} {\bibinfo  {journal}
  {Chaos An Interdiscip. J. Nonlinear Sci.}\ }\textbf {\bibinfo {volume}
  {25}},\ \bibinfo {pages} {043109} (\bibinfo {year} {2015})}\BibitemShut
  {NoStop}%
\bibitem [{\citenamefont {Fukushima}\ and\ \citenamefont
  {Yamada}(1995)}]{Fukushima_PLA95}%
  \BibitemOpen
  \bibfield  {author} {\bibinfo {author} {\bibfnamefont {K.}~\bibnamefont
  {Fukushima}}\ and\ \bibinfo {author} {\bibfnamefont {T.}~\bibnamefont
  {Yamada}},\ }\href {\doibase 10.1016/0375-9601(95)00175-3} {\bibfield
  {journal} {\bibinfo  {journal} {Phys. Lett. A}\ }\textbf {\bibinfo {volume}
  {200}},\ \bibinfo {pages} {350} (\bibinfo {year} {1995})}\BibitemShut
  {NoStop}%
\bibitem [{\citenamefont {Higuchi}\ \emph {et~al.}(1998)\citenamefont
  {Higuchi}, \citenamefont {Fukushima},\ and\ \citenamefont
  {Yamada}}]{Higuchi_CSF98}%
  \BibitemOpen
  \bibfield  {author} {\bibinfo {author} {\bibfnamefont {M.}~\bibnamefont
  {Higuchi}}, \bibinfo {author} {\bibfnamefont {K.}~\bibnamefont {Fukushima}},
  \ and\ \bibinfo {author} {\bibfnamefont {T.}~\bibnamefont {Yamada}},\ }\href
  {\doibase 10.1016/S0960-0779(97)00081-7} {\bibfield  {journal} {\bibinfo
  {journal} {Chaos, Solitons {\&} Fractals}\ }\textbf {\bibinfo {volume} {9}},\
  \bibinfo {pages} {845} (\bibinfo {year} {1998})}\BibitemShut {NoStop}%
\bibitem [{\citenamefont {Dmitriev}\ \emph {et~al.}(2001)\citenamefont
  {Dmitriev}, \citenamefont {Kivshar},\ and\ \citenamefont
  {Shigenari}}]{Dmitriev_PRE01}%
  \BibitemOpen
  \bibfield  {author} {\bibinfo {author} {\bibfnamefont {S.~V.}\ \bibnamefont
  {Dmitriev}}, \bibinfo {author} {\bibfnamefont {Y.~S.}\ \bibnamefont
  {Kivshar}}, \ and\ \bibinfo {author} {\bibfnamefont {T.}~\bibnamefont
  {Shigenari}},\ }\href {\doibase 10.1103/PhysRevE.64.056613} {\bibfield
  {journal} {\bibinfo  {journal} {Phys. Rev. E}\ }\textbf {\bibinfo {volume}
  {64}},\ \bibinfo {pages} {056613} (\bibinfo {year} {2001})}\BibitemShut
  {NoStop}%
\bibitem [{\citenamefont {Dmitriev}\ \emph {et~al.}(2002)\citenamefont
  {Dmitriev}, \citenamefont {Kivshar},\ and\ \citenamefont
  {Shigenari}}]{Dmitriev_PB02}%
  \BibitemOpen
  \bibfield  {author} {\bibinfo {author} {\bibfnamefont {S.~V.}\ \bibnamefont
  {Dmitriev}}, \bibinfo {author} {\bibfnamefont {Y.~S.}\ \bibnamefont
  {Kivshar}}, \ and\ \bibinfo {author} {\bibfnamefont {T.}~\bibnamefont
  {Shigenari}},\ }\href {\doibase 10.1016/S0921-4526(02)00442-8} {\bibfield
  {journal} {\bibinfo  {journal} {Phys. B Condens. Matter}\ }\textbf {\bibinfo
  {volume} {316-317}},\ \bibinfo {pages} {139} (\bibinfo {year}
  {2002})}\BibitemShut {NoStop}%
\bibitem [{\citenamefont {Dmitriev}\ \emph {et~al.}(2008)\citenamefont
  {Dmitriev}, \citenamefont {Kevrekidis},\ and\ \citenamefont
  {Kivshar}}]{Dmitriev_PRE08}%
  \BibitemOpen
  \bibfield  {author} {\bibinfo {author} {\bibfnamefont {S.~V.}\ \bibnamefont
  {Dmitriev}}, \bibinfo {author} {\bibfnamefont {P.~G.}\ \bibnamefont
  {Kevrekidis}}, \ and\ \bibinfo {author} {\bibfnamefont {Y.~S.}\ \bibnamefont
  {Kivshar}},\ }\href {\doibase 10.1103/PhysRevE.78.046604} {\bibfield
  {journal} {\bibinfo  {journal} {Phys. Rev. E}\ }\textbf {\bibinfo {volume}
  {78}},\ \bibinfo {pages} {046604} (\bibinfo {year} {2008})}\BibitemShut
  {NoStop}%
\bibitem [{\citenamefont {Myatt}\ \emph {et~al.}(1997)\citenamefont {Myatt},
  \citenamefont {Burt}, \citenamefont {Ghrist}, \citenamefont {Cornell},\ and\
  \citenamefont {Wieman}}]{Myatt_PRL97}%
  \BibitemOpen
  \bibfield  {author} {\bibinfo {author} {\bibfnamefont {C.~J.}\ \bibnamefont
  {Myatt}}, \bibinfo {author} {\bibfnamefont {E.~A.}\ \bibnamefont {Burt}},
  \bibinfo {author} {\bibfnamefont {R.~W.}\ \bibnamefont {Ghrist}}, \bibinfo
  {author} {\bibfnamefont {E.~A.}\ \bibnamefont {Cornell}}, \ and\ \bibinfo
  {author} {\bibfnamefont {C.~E.}\ \bibnamefont {Wieman}},\ }\href {\doibase
  10.1103/PhysRevLett.78.586} {\bibfield  {journal} {\bibinfo  {journal} {Phys.
  Rev. Lett.}\ }\textbf {\bibinfo {volume} {78}},\ \bibinfo {pages} {586}
  (\bibinfo {year} {1997})}\BibitemShut {NoStop}%
\bibitem [{\citenamefont {Stamper-Kurn}\ \emph {et~al.}(1998)\citenamefont
  {Stamper-Kurn}, \citenamefont {Andrews}, \citenamefont {Chikkatur},
  \citenamefont {Inouye}, \citenamefont {Miesner}, \citenamefont {Stenger},\
  and\ \citenamefont {Ketterle}}]{Stamper-Kurn_PRL98}%
  \BibitemOpen
  \bibfield  {author} {\bibinfo {author} {\bibfnamefont {D.~M.}\ \bibnamefont
  {Stamper-Kurn}}, \bibinfo {author} {\bibfnamefont {M.~R.}\ \bibnamefont
  {Andrews}}, \bibinfo {author} {\bibfnamefont {A.~P.}\ \bibnamefont
  {Chikkatur}}, \bibinfo {author} {\bibfnamefont {S.}~\bibnamefont {Inouye}},
  \bibinfo {author} {\bibfnamefont {H.-J.}\ \bibnamefont {Miesner}}, \bibinfo
  {author} {\bibfnamefont {J.}~\bibnamefont {Stenger}}, \ and\ \bibinfo
  {author} {\bibfnamefont {W.}~\bibnamefont {Ketterle}},\ }\href {\doibase
  10.1103/PhysRevLett.80.2027} {\bibfield  {journal} {\bibinfo  {journal}
  {Phys. Rev. Lett.}\ }\textbf {\bibinfo {volume} {80}},\ \bibinfo {pages}
  {2027} (\bibinfo {year} {1998})}\BibitemShut {NoStop}%
\bibitem [{\citenamefont {Cardoso}\ \emph {et~al.}(2012)\citenamefont
  {Cardoso}, \citenamefont {Avelar},\ and\ \citenamefont
  {Bazeia}}]{Cardoso_PRE12}%
  \BibitemOpen
  \bibfield  {author} {\bibinfo {author} {\bibfnamefont {W.~B.}\ \bibnamefont
  {Cardoso}}, \bibinfo {author} {\bibfnamefont {A.~T.}\ \bibnamefont {Avelar}},
  \ and\ \bibinfo {author} {\bibfnamefont {D.}~\bibnamefont {Bazeia}},\ }\href
  {\doibase 10.1103/PhysRevE.86.027601} {\bibfield  {journal} {\bibinfo
  {journal} {Phys. Rev. E}\ }\textbf {\bibinfo {volume} {86}},\ \bibinfo
  {pages} {27601} (\bibinfo {year} {2012})}\BibitemShut {NoStop}%
\bibitem [{\citenamefont {Cardoso}\ \emph {et~al.}(2010)\citenamefont
  {Cardoso}, \citenamefont {Avelar}, \citenamefont {Bazeia},\ and\
  \citenamefont {Hussein}}]{Cardoso_PLA10-2}%
  \BibitemOpen
  \bibfield  {author} {\bibinfo {author} {\bibfnamefont {W.~B.}\ \bibnamefont
  {Cardoso}}, \bibinfo {author} {\bibfnamefont {A.~T.}\ \bibnamefont {Avelar}},
  \bibinfo {author} {\bibfnamefont {D.}~\bibnamefont {Bazeia}}, \ and\ \bibinfo
  {author} {\bibfnamefont {M.~S.}\ \bibnamefont {Hussein}},\ }\href {\doibase
  10.1016/j.physleta.2010.03.065} {\bibfield  {journal} {\bibinfo  {journal}
  {Phys. Lett. A}\ }\textbf {\bibinfo {volume} {374}},\ \bibinfo {pages} {2356}
  (\bibinfo {year} {2010})}\BibitemShut {NoStop}%
\bibitem [{\citenamefont {Kogelnik}\ and\ \citenamefont
  {Schmidt}(1976)}]{Kogelnik_IEEE76}%
  \BibitemOpen
  \bibfield  {author} {\bibinfo {author} {\bibfnamefont {H.}~\bibnamefont
  {Kogelnik}}\ and\ \bibinfo {author} {\bibfnamefont {R.}~\bibnamefont
  {Schmidt}},\ }\href {\doibase 10.1109/JQE.1976.1069190} {\bibfield  {journal}
  {\bibinfo  {journal} {IEEE J. Quantum Electron.}\ }\textbf {\bibinfo {volume}
  {12}},\ \bibinfo {pages} {396} (\bibinfo {year} {1976})}\BibitemShut
  {NoStop}%
\bibitem [{\citenamefont {Bergh}\ \emph {et~al.}(1980)\citenamefont {Bergh},
  \citenamefont {Kotler},\ and\ \citenamefont {Shaw}}]{Bergh_EL80}%
  \BibitemOpen
  \bibfield  {author} {\bibinfo {author} {\bibfnamefont {R.}~\bibnamefont
  {Bergh}}, \bibinfo {author} {\bibfnamefont {G.}~\bibnamefont {Kotler}}, \
  and\ \bibinfo {author} {\bibfnamefont {H.}~\bibnamefont {Shaw}},\ }\href
  {\doibase 10.1049/el:19800191} {\bibfield  {journal} {\bibinfo  {journal}
  {Electron. Lett.}\ }\textbf {\bibinfo {volume} {16}},\ \bibinfo {pages} {260}
  (\bibinfo {year} {1980})}\BibitemShut {NoStop}%
\bibitem [{\citenamefont {Streltsov}\ and\ \citenamefont
  {Borrelli}(2001)}]{Streltsov_OL01}%
  \BibitemOpen
  \bibfield  {author} {\bibinfo {author} {\bibfnamefont {A.~M.}\ \bibnamefont
  {Streltsov}}\ and\ \bibinfo {author} {\bibfnamefont {N.~F.}\ \bibnamefont
  {Borrelli}},\ }\href {\doibase 10.1364/OL.26.000042} {\bibfield  {journal}
  {\bibinfo  {journal} {Opt. Lett.}\ }\textbf {\bibinfo {volume} {26}},\
  \bibinfo {pages} {42} (\bibinfo {year} {2001})}\BibitemShut {NoStop}%
\bibitem [{\citenamefont {Alves}\ \emph {et~al.}(2015)\citenamefont {Alves},
  \citenamefont {Cardoso},\ and\ \citenamefont {Avelar}}]{Alves_arXiv15}%
  \BibitemOpen
  \bibfield  {author} {\bibinfo {author} {\bibfnamefont {E.~O.}\ \bibnamefont
  {Alves}}, \bibinfo {author} {\bibfnamefont {W.~B.}\ \bibnamefont {Cardoso}},
  \ and\ \bibinfo {author} {\bibfnamefont {A.~T.}\ \bibnamefont {Avelar}},\
  }\href {http://arxiv.org/abs/1505.06719} {\  (\bibinfo {year} {2015})},\
  \Eprint {http://arxiv.org/abs/1505.06719} {arXiv:1505.06719} \BibitemShut
  {NoStop}%
\bibitem [{\citenamefont {Biswas}\ and\ \citenamefont
  {Konar}(2006)}]{Biswas_06}%
  \BibitemOpen
  \bibfield  {author} {\bibinfo {author} {\bibfnamefont {A.}~\bibnamefont
  {Biswas}}\ and\ \bibinfo {author} {\bibfnamefont {S.}~\bibnamefont {Konar}},\
  }\href {https://books.google.com.br/books?id=WerVB246wgsC} {\emph {\bibinfo
  {title} {{Introduction to non-Kerr Law Optical Solitons}}}},\ Chapman {\&}
  Hall/CRC Applied Mathematics {\&} Nonlinear Science\ (\bibinfo  {publisher}
  {CRC Press},\ \bibinfo {year} {2006})\BibitemShut {NoStop}%
\bibitem [{\citenamefont {Hern{\'{a}}ndez}\ and\ \citenamefont
  {Remaud}(1981)}]{Hernandez_PA81}%
  \BibitemOpen
  \bibfield  {author} {\bibinfo {author} {\bibfnamefont {E.~S.}\ \bibnamefont
  {Hern{\'{a}}ndez}}\ and\ \bibinfo {author} {\bibfnamefont {B.}~\bibnamefont
  {Remaud}},\ }\href {\doibase 10.1016/0378-4371(81)90066-2} {\bibfield
  {journal} {\bibinfo  {journal} {Phys. A Stat. Mech. its Appl.}\ }\textbf
  {\bibinfo {volume} {105}},\ \bibinfo {pages} {130} (\bibinfo {year}
  {1981})}\BibitemShut {NoStop}%
\bibitem [{\citenamefont {Hefter}(1985)}]{Hefter_PRA85}%
  \BibitemOpen
  \bibfield  {author} {\bibinfo {author} {\bibfnamefont {E.~F.}\ \bibnamefont
  {Hefter}},\ }\href {\doibase 10.1103/PhysRevA.32.1201} {\bibfield  {journal}
  {\bibinfo  {journal} {Phys. Rev. A}\ }\textbf {\bibinfo {volume} {32}},\
  \bibinfo {pages} {1201} (\bibinfo {year} {1985})}\BibitemShut {NoStop}%
\bibitem [{\citenamefont {Kr{\'{o}}likowski}\ \emph {et~al.}(2000)\citenamefont
  {Kr{\'{o}}likowski}, \citenamefont {Edmundson},\ and\ \citenamefont
  {Bang}}]{Krolikowski_PRE00}%
  \BibitemOpen
  \bibfield  {author} {\bibinfo {author} {\bibfnamefont {W.}~\bibnamefont
  {Kr{\'{o}}likowski}}, \bibinfo {author} {\bibfnamefont {D.}~\bibnamefont
  {Edmundson}}, \ and\ \bibinfo {author} {\bibfnamefont {O.}~\bibnamefont
  {Bang}},\ }\href {\doibase 10.1103/PhysRevE.61.3122} {\bibfield  {journal}
  {\bibinfo  {journal} {Phys. Rev. E}\ }\textbf {\bibinfo {volume} {61}},\
  \bibinfo {pages} {3122} (\bibinfo {year} {2000})}\BibitemShut {NoStop}%
\bibitem [{\citenamefont {Buljan}\ \emph {et~al.}(2003)\citenamefont {Buljan},
  \citenamefont {{\v{S}}iber}, \citenamefont {Solja{\v{c}}i{\'{c}}},
  \citenamefont {Schwartz}, \citenamefont {Segev},\ and\ \citenamefont
  {Christodoulides}}]{Buljan_PRE03}%
  \BibitemOpen
  \bibfield  {author} {\bibinfo {author} {\bibfnamefont {H.}~\bibnamefont
  {Buljan}}, \bibinfo {author} {\bibfnamefont {A.}~\bibnamefont {{\v{S}}iber}},
  \bibinfo {author} {\bibfnamefont {M.}~\bibnamefont {Solja{\v{c}}i{\'{c}}}},
  \bibinfo {author} {\bibfnamefont {T.}~\bibnamefont {Schwartz}}, \bibinfo
  {author} {\bibfnamefont {M.}~\bibnamefont {Segev}}, \ and\ \bibinfo {author}
  {\bibfnamefont {D.~N.}\ \bibnamefont {Christodoulides}},\ }\href {\doibase
  10.1103/PhysRevE.68.036607} {\bibfield  {journal} {\bibinfo  {journal} {Phys.
  Rev. E}\ }\textbf {\bibinfo {volume} {68}},\ \bibinfo {pages} {036607}
  (\bibinfo {year} {2003})}\BibitemShut {NoStop}%
\bibitem [{\citenamefont {{De Martino}}\ and\ \citenamefont
  {Lauro}(2004)}]{DeMartino_WASCOM04}%
  \BibitemOpen
  \bibfield  {author} {\bibinfo {author} {\bibfnamefont {S.}~\bibnamefont {{De
  Martino}}}\ and\ \bibinfo {author} {\bibfnamefont {G.}~\bibnamefont
  {Lauro}},\ }in\ \href {\doibase 10.1142/9789812702937{\_}0019} {\emph
  {\bibinfo {booktitle} {Waves Stab. Contin. Media}}},\ \bibinfo {editor}
  {edited by\ \bibinfo {editor} {\bibfnamefont {R.}~\bibnamefont {Monaco}}}\
  (\bibinfo  {publisher} {WORLD SCIENTIFIC},\ \bibinfo {year} {2004})\ pp.\
  \bibinfo {pages} {148--152}\BibitemShut {NoStop}%
\bibitem [{\citenamefont {Martino}\ \emph {et~al.}(2003)\citenamefont
  {Martino}, \citenamefont {Falanga}, \citenamefont {Godano},\ and\
  \citenamefont {Lauro}}]{Martino_EPL03}%
  \BibitemOpen
  \bibfield  {author} {\bibinfo {author} {\bibfnamefont {S.~D.}\ \bibnamefont
  {Martino}}, \bibinfo {author} {\bibfnamefont {M.}~\bibnamefont {Falanga}},
  \bibinfo {author} {\bibfnamefont {C.}~\bibnamefont {Godano}}, \ and\ \bibinfo
  {author} {\bibfnamefont {G.}~\bibnamefont {Lauro}},\ }\href {\doibase
  10.1209/epl/i2003-00547-6} {\bibfield  {journal} {\bibinfo  {journal}
  {Europhys. Lett.}\ }\textbf {\bibinfo {volume} {63}},\ \bibinfo {pages} {472}
  (\bibinfo {year} {2003})}\BibitemShut {NoStop}%
\bibitem [{\citenamefont {Biswas}\ and\ \citenamefont
  {Milovi{\'{c}}}(2010)}]{Biswas_CNSNS10}%
  \BibitemOpen
  \bibfield  {author} {\bibinfo {author} {\bibfnamefont {A.}~\bibnamefont
  {Biswas}}\ and\ \bibinfo {author} {\bibfnamefont {D.}~\bibnamefont
  {Milovi{\'{c}}}},\ }\href {\doibase 10.1016/j.cnsns.2010.01.022} {\bibfield
  {journal} {\bibinfo  {journal} {Commun. Nonlinear Sci. Numer. Simul.}\
  }\textbf {\bibinfo {volume} {15}},\ \bibinfo {pages} {3763} (\bibinfo {year}
  {2010})}\BibitemShut {NoStop}%
\bibitem [{\citenamefont {Biswas}\ \emph {et~al.}(2012)\citenamefont {Biswas},
  \citenamefont {Fessak}, \citenamefont {Johnson}, \citenamefont {Beatrice},
  \citenamefont {Milovic}, \citenamefont {Jovanoski}, \citenamefont {Kohl},\
  and\ \citenamefont {Majid}}]{Biswas_OLT12}%
  \BibitemOpen
  \bibfield  {author} {\bibinfo {author} {\bibfnamefont {A.}~\bibnamefont
  {Biswas}}, \bibinfo {author} {\bibfnamefont {M.}~\bibnamefont {Fessak}},
  \bibinfo {author} {\bibfnamefont {S.}~\bibnamefont {Johnson}}, \bibinfo
  {author} {\bibfnamefont {S.}~\bibnamefont {Beatrice}}, \bibinfo {author}
  {\bibfnamefont {D.}~\bibnamefont {Milovic}}, \bibinfo {author} {\bibfnamefont
  {Z.}~\bibnamefont {Jovanoski}}, \bibinfo {author} {\bibfnamefont
  {R.}~\bibnamefont {Kohl}}, \ and\ \bibinfo {author} {\bibfnamefont
  {F.}~\bibnamefont {Majid}},\ }\href {\doibase
  10.1016/j.optlastec.2011.07.001} {\bibfield  {journal} {\bibinfo  {journal}
  {Opt. Laser Technol.}\ }\textbf {\bibinfo {volume} {44}},\ \bibinfo {pages}
  {263} (\bibinfo {year} {2012})}\BibitemShut {NoStop}%
\bibitem [{\citenamefont {Zhou}\ \emph {et~al.}(2013)\citenamefont {Zhou},
  \citenamefont {Yao}, \citenamefont {Xu},\ and\ \citenamefont
  {Liu}}]{Zhou_Optik13}%
  \BibitemOpen
  \bibfield  {author} {\bibinfo {author} {\bibfnamefont {Q.}~\bibnamefont
  {Zhou}}, \bibinfo {author} {\bibfnamefont {D.}~\bibnamefont {Yao}}, \bibinfo
  {author} {\bibfnamefont {Q.}~\bibnamefont {Xu}}, \ and\ \bibinfo {author}
  {\bibfnamefont {X.}~\bibnamefont {Liu}},\ }\href {\doibase
  10.1016/j.ijleo.2012.07.045} {\bibfield  {journal} {\bibinfo  {journal} {Opt.
  - Int. J. Light Electron Opt.}\ }\textbf {\bibinfo {volume} {124}},\ \bibinfo
  {pages} {2368} (\bibinfo {year} {2013})}\BibitemShut {NoStop}%
\bibitem [{\citenamefont {Hilal}\ \emph {et~al.}(2014)\citenamefont {Hilal},
  \citenamefont {Alshaery}, \citenamefont {Bhrawy}, \citenamefont {Bhosale},\
  and\ \citenamefont {Biswas}}]{Hilal_Optik14}%
  \BibitemOpen
  \bibfield  {author} {\bibinfo {author} {\bibfnamefont {E.~M.}\ \bibnamefont
  {Hilal}}, \bibinfo {author} {\bibfnamefont {A.~A.}\ \bibnamefont {Alshaery}},
  \bibinfo {author} {\bibfnamefont {A.~H.}\ \bibnamefont {Bhrawy}}, \bibinfo
  {author} {\bibfnamefont {B.}~\bibnamefont {Bhosale}}, \ and\ \bibinfo
  {author} {\bibfnamefont {A.}~\bibnamefont {Biswas}},\ }\href {\doibase
  10.1016/j.ijleo.2014.05.041} {\bibfield  {journal} {\bibinfo  {journal} {Opt.
  - Int. J. Light Electron Opt.}\ }\textbf {\bibinfo {volume} {125}},\ \bibinfo
  {pages} {4589} (\bibinfo {year} {2014})}\BibitemShut {NoStop}%
\bibitem [{\citenamefont {Cala{\c{c}}a}\ \emph {et~al.}(2014)\citenamefont
  {Cala{\c{c}}a}, \citenamefont {Avelar}, \citenamefont {Bazeia},\ and\
  \citenamefont {Cardoso}}]{Calaca_CNSNS14}%
  \BibitemOpen
  \bibfield  {author} {\bibinfo {author} {\bibfnamefont {L.}~\bibnamefont
  {Cala{\c{c}}a}}, \bibinfo {author} {\bibfnamefont {A.~T.}\ \bibnamefont
  {Avelar}}, \bibinfo {author} {\bibfnamefont {D.}~\bibnamefont {Bazeia}}, \
  and\ \bibinfo {author} {\bibfnamefont {W.~B.}\ \bibnamefont {Cardoso}},\
  }\href {\doibase 10.1016/j.cnsns.2014.02.002} {\bibfield  {journal} {\bibinfo
   {journal} {Commun. Nonlinear Sci. Numer. Simul.}\ }\textbf {\bibinfo
  {volume} {19}},\ \bibinfo {pages} {2928} (\bibinfo {year}
  {2014})}\BibitemShut {NoStop}%
\bibitem [{\citenamefont {Biswas}\ \emph {et~al.}(2010)\citenamefont {Biswas},
  \citenamefont {Cleary}, \citenamefont {Watson},\ and\ \citenamefont
  {Milovic}}]{Biswas_AMC10}%
  \BibitemOpen
  \bibfield  {author} {\bibinfo {author} {\bibfnamefont {A.}~\bibnamefont
  {Biswas}}, \bibinfo {author} {\bibfnamefont {C.}~\bibnamefont {Cleary}},
  \bibinfo {author} {\bibfnamefont {J.~E.}\ \bibnamefont {Watson}}, \ and\
  \bibinfo {author} {\bibfnamefont {D.}~\bibnamefont {Milovic}},\ }\href
  {\doibase 10.1016/j.amc.2010.07.032} {\bibfield  {journal} {\bibinfo
  {journal} {Appl. Math. Comput.}\ }\textbf {\bibinfo {volume} {217}},\
  \bibinfo {pages} {2891} (\bibinfo {year} {2010})}\BibitemShut {NoStop}%
\bibitem [{\citenamefont {Biswas}\ \emph {et~al.}(2011)\citenamefont {Biswas},
  \citenamefont {Topkara}, \citenamefont {Johnson}, \citenamefont {Zerrad},\
  and\ \citenamefont {Konar}}]{BISWAS_JNOPM11}%
  \BibitemOpen
  \bibfield  {author} {\bibinfo {author} {\bibfnamefont {A.}~\bibnamefont
  {Biswas}}, \bibinfo {author} {\bibfnamefont {E.}~\bibnamefont {Topkara}},
  \bibinfo {author} {\bibfnamefont {S.}~\bibnamefont {Johnson}}, \bibinfo
  {author} {\bibfnamefont {E.}~\bibnamefont {Zerrad}}, \ and\ \bibinfo {author}
  {\bibfnamefont {S.}~\bibnamefont {Konar}},\ }\href {\doibase
  10.1142/S0218863511006108} {\bibfield  {journal} {\bibinfo  {journal} {J.
  Nonlinear Opt. Phys. Mater.}\ }\textbf {\bibinfo {volume} {20}},\ \bibinfo
  {pages} {309} (\bibinfo {year} {2011})}\BibitemShut {NoStop}%
\bibitem [{\citenamefont {Yang}(2010)}]{Yang_10}%
  \BibitemOpen
  \bibfield  {author} {\bibinfo {author} {\bibfnamefont {J.}~\bibnamefont
  {Yang}},\ }\href {\doibase 10.1137/1.9780898719680} {\emph {\bibinfo {title}
  {{Nonlinear Waves in Integrable and Nonintegrable Systems}}}}\ (\bibinfo
  {publisher} {Society for Industrial and Applied Mathematics},\ \bibinfo
  {year} {2010})\BibitemShut {NoStop}%
\bibitem [{\citenamefont {Campbell}\ \emph {et~al.}(1983)\citenamefont
  {Campbell}, \citenamefont {Schonfeld},\ and\ \citenamefont
  {Wingate}}]{Campbell_PD83}%
  \BibitemOpen
  \bibfield  {author} {\bibinfo {author} {\bibfnamefont {D.~K.}\ \bibnamefont
  {Campbell}}, \bibinfo {author} {\bibfnamefont {J.~F.}\ \bibnamefont
  {Schonfeld}}, \ and\ \bibinfo {author} {\bibfnamefont {C.~A.}\ \bibnamefont
  {Wingate}},\ }\href {\doibase 10.1016/0167-2789(83)90289-0} {\bibfield
  {journal} {\bibinfo  {journal} {Phys. D Nonlinear Phenom.}\ }\textbf
  {\bibinfo {volume} {9}},\ \bibinfo {pages} {1} (\bibinfo {year}
  {1983})}\BibitemShut {NoStop}%
\bibitem [{\citenamefont {Anninos}\ \emph {et~al.}(1991)\citenamefont
  {Anninos}, \citenamefont {Oliveira},\ and\ \citenamefont
  {Matzner}}]{Anninos_PRD91}%
  \BibitemOpen
  \bibfield  {author} {\bibinfo {author} {\bibfnamefont {P.}~\bibnamefont
  {Anninos}}, \bibinfo {author} {\bibfnamefont {S.}~\bibnamefont {Oliveira}}, \
  and\ \bibinfo {author} {\bibfnamefont {R.~A.}\ \bibnamefont {Matzner}},\
  }\href {\doibase 10.1103/PhysRevD.44.1147} {\bibfield  {journal} {\bibinfo
  {journal} {Phys. Rev. D}\ }\textbf {\bibinfo {volume} {44}},\ \bibinfo
  {pages} {1147} (\bibinfo {year} {1991})}\BibitemShut {NoStop}%
\end{thebibliography}%

\end{document}